\documentclass[times,twocolumn]{aastex62}
\usepackage[english]{babel}
\usepackage[utf8x]{inputenc}
\usepackage[T1]{fontenc}
\usepackage{amssymb}
\usepackage{multirow}
\usepackage{natbib}
\usepackage{afterpage}
\usepackage[abs]{overpic}

\def\bi#1{\hbox{\boldmath{$#1$}}}

\newcommand{\mpc}{\ensuremath{\, h^{-1}\,\mathrm{Mpc}\, }}

\newcommand{\lya}{Ly$\alpha$}

\newcommand{\beq}{\begin{equation}}
\newcommand{\eeq}{\end{equation}}
\newcommand{\bc}{\begin{center}}
\newcommand{\ec}{\end{center}}
\newcommand{\bfig}{\begin{figure}}
\newcommand{\efig}{\end{figure}}

\usepackage{amsmath}
\usepackage{graphicx}
\usepackage[colorinlistoftodos]{todonotes}
\usepackage[colorlinks=true]{hyperref}
\usepackage{comment}

\begin{document}

\title{TARDIS Paper I: A Constrained Reconstruction Approach to \\ Modeling the $z\sim 2.5$ Cosmic Web Probed by Lyman-$\alpha$
Forest Tomography}
\author{Benjamin Horowitz}
\affil{Lawrence Berkeley National Lab, 1 Cyclotron Road, Berkeley, CA 94720, USA}
\affil{Department of Physics, University of California at Berkeley, Berkeley, CA 94720, USA}
\email{bhorowitz@berkeley.edu}
\affil{Kavli IPMU (WPI), UTIAS, The University of Tokyo, Kashiwa, Chiba 277-8583, Japan}

\author{Khee-Gan Lee}
\affil{Kavli IPMU (WPI), UTIAS, The University of Tokyo, Kashiwa, Chiba 277-8583, Japan}
\affil{Lawrence Berkeley National Lab, 1 Cyclotron Road, Berkeley, CA 94720, USA}

\author{Martin White}
\affil{Lawrence Berkeley National Lab, 1 Cyclotron Road, Berkeley, CA 94720, USA}
\affil{Department of Physics, University of California at Berkeley, Berkeley, CA 94720, USA}
\affil{Department of Astronomy, University of California at Berkeley, Berkeley, CA 94720, USA}
\author{Alex Krolewski}
\affil{Department of Astronomy, University of California at Berkeley, Berkeley, CA 94720, USA}
\author{Metin Ata}
\affil{Kavli IPMU (WPI), UTIAS, The University of Tokyo, Kashiwa, Chiba 277-8583, Japan}

\begin{abstract}

Recent Lyman-$\alpha$ forest tomography measurements of the intergalactic medium (IGM) have revealed a wealth of cosmic structures at high redshift ($z\sim 2.5$). In this work, we present the Tomographic Absorption Reconstruction and Density Inference Scheme (TARDIS), a new chrono-cosmographic analysis tool for understanding the formation and evolution of these observed structures. We use maximum likelihood techniques with a fast non-linear gravitational model to reconstruct the initial density field of the observed regions. We find that TARDIS allows accurate reconstruction of smaller scale structures than standard Wiener filtering techniques.
Applying this technique to mock Lyman-$\alpha$ forest data sets that simulate ongoing and future surveys such as CLAMATO, Subaru-PFS or the ELTs, 
we are able to infer the underlying matter density field at observed redshift 
and classify the cosmic web structures. We find good agreement with the underlying truth both in the characteristic eigenvalues and eigenvectors of the
pseudo-deformation tensor, with the eigenvalues inferred from 30m-class telescopes correlated at $r=0.95$ relative to the truth. As an output of this method, we are able to further evolve the inferred structures to late time ($z=0$), and also track the trajectories of coeval $z=2.5$ galaxies to their $z=0$ cosmic web environments. \end{abstract}
\

\section{Introduction}

A major goal of modern astrophysics is to understand how galaxies form and evolve from initial density fluctuations to the current day. Over the past few decades it has become increasingly clear that the surrounding large scale structures around galaxies play a critical role in their formation, morphology, and evolution \citep{1980Dressler,2004Kauffmann}. There has also been new theoretical understanding on how these large scale dark matter structures evolve, from both an analytical approach and from numerical simulations 
\bibpunct[]{(}{)}{,}{a}{}{;}
\citep[see][ for an overview]{reviewGal2}.
\bibpunct[; ]{(}{)}{,}{a}{}{,}
However, our understanding of the small scale processes driving galaxy evolution remains poor, with many competing models \citep{reviewGal2,reviewGal}. Part of the challenge lies in the fact that most observations linking galaxy evolution and large-scale structure are at
low redshifts, whereas most of the galaxy- and star-formation in the Universe peaked at the so-called `Cosmic Noon' epoch at $z\sim 1.5-3$ \citep{cosmicnoon} which 
remain out of reach of most large-scale structure surveys. 

There are many indications of the interconnected nature of cosmic structure and galactic evolution at high redshift. Numerous studies have found that low-redshift galaxies living in cluster environments have lower star formation rates and significantly older stellar ages than those in the field \citep{2005Wake,2009Skibba}. This indicates that these regions underwent significant star formation and quenching at high redshift ($z>1.5$) \citep{2010Tran}. This is further supported by simulation work showing that protoclusters produce roughly half of their stellar content at $2<z<4$ and are therefore an important contribution to the overall cosmic star formation rate \citep{2017Chiang}.
Beyond protoclusters, there is evidence to suggest that star formation proprieties may further depend on where the galaxy is first formed in the cluster or falls in along filamentary structure \citep{2008Porter}. Similarly, hydrodynamical simulations \citep{2014Dubois} have suggested that the spin of galaxies may depend on the filament orientation, with simulated red and blue galaxies aligning perpendicular and parallel to the filament, respectively. Very little data is available tracing these cosmic structures at high redshift, but next generation surveys will provide the depth over sufficient sky coverage to better constrain these astrophysical processes \citep{2019Kartaltepe,2019Overzier}.

Understanding these complex relationships between baryonic properties and dark matter in the context of the overall large-scale structure environment is not only useful in modeling galaxy formation, but is also crucial in exploiting galaxies as biased tracers of large scale structure for cosmological constraints \citep{galaxybias}. The relationships between cosmic web structures and bias has been explored in the case of tidal shear bias \citep{2012Baldauf}
and more recently in the case of assembly bias \citep{2019Ramakrishnan}. Quantifying the sources of bias will be needed when extending galaxy clustering surveys into the nonlinear regime where the particulars of the cosmic web may play a role \citep{2019Alam}, or in cosmic shear surveys where intrinsic alignments of galaxies will contribute substantial systematic uncertainty to precision cosmological measurements \citep{2015Troxel}.

So far most studies of the cosmic web have used optically selected galaxies from spectrosopic redshift surveys as a tracer of the cosmic web. As a high number density
(and threfore high spectroscopic sampling rate) is necessary for this sort of survey, this technique becomes increasingly expensive at higher redshift.
 The current state-of-the-art galaxy survey probing the high-redshift cosmic web is the VIPERS survey \citep{guzzo:2014} on the {\it Very Large Telescope} (VLT),
which has obtained redshifts for 100,000 galaxies over 24 deg$^2$ as the largest-ever spectroscopic campaign on that facility. 
This enabled a cosmic web analysis in the redshift range $0.4<z<1.0$ \citep{malavasi:2016}, which suggested segregation of massive galaxies
towards filaments already at this redshift. Over the next few years, new massively-multiplexed fiber spectrographs on 8m-class telescopes, such as VLT-MOONS \citep{moons} and Subaru-PFS \citep[PFS;][]{subaru}, will allow 
such high-sampling rate galaxy surveys to push to $z\sim 1.5$, but would be prohibitively expensive at the ``Cosmic Noon'' epoch of $z\sim 2-3$. 

In recent years, however, ``intergalactic medium (IGM) tomography'' \citep{2001Pichon,caucci,Lee2014Theory,Stark2015} of the hydrogen \lya\ forest 
provides a complementary approach to mapping high-redshift large-scale structure. This technique uses dense configurations of closely-spaced
star-forming galaxies, in addition to quasars, as background sources to probe 
the three-dimensional (3D) structure of the optically thin IGM gas at $z>2$ on scales of
several comoving Mpc. 
The ongoing COSMOS Lyman Alpha Mapping And 
Tomographic Observations (CLAMATO) survey is the first observational program to 
implement IGM tomography, and now has 240 sightlines covering a $\sim 600$ square arcmin footprint within 
the COSMOS field, yielding a 3D tomographic map of the $2.05<z<2.55$ \lya\ forest \citep{Lee2017}. A number of $z\sim 2.3$ 
cosmic structures already have been detected in the CLAMATO data, including protoclusters \citep{2016LeeColossus} and cosmic voids \citep{2018Krolewski}. 

In the coming years, a number of next generation spectroscopic surveys will radically increase the observational resources available for IGM tomgraphy, including the Subaru Prime Focus Spectrograph and Maunakea Spectroscopic Explorer
\citep[MSE;][]{2016MSE}. These telescopes will offer multiplex factors of several thousand over $\sim $ 1 deg$^2$ fields of view, allowing several times the 
volume of the current CLAMATO data to be observed within a single night.
Meanwhile, with far sparser sightline number density but significantly larger sky coverage, the Dark Energy Spectroscopic Instrument \citep[DESI;][]{desi} could be another interesting platform for Lyman-$\alpha$ forest tomography to probe large-scale over-densities.
% particularly when combined with its overlapping galaxy sample. (there is no overlapping galaxy sample in DESI)
Farther into the future, the thirty-meter class facilities such as
Thirty Meter Telescope \citep[TMT;][]{2015TMT}, Giant Magellan Telescope \citep[GMT;][]{2012GMT}, and 
European Extremely Large Telescope \citep[EELT;][]{2014EELT}, will have smaller fields-of-view but dramatically
improved sensitivity for faint background sources at much greater sightline densities
that can probe spatial scales of $\sim 1$ cMpc and below.
The need for accurate modeling of the formation and evolution of galaxies and galaxy clusters increases in order to maximize the science return of these facilities.

The current standard procedure for IGM tomography analysis is to create a Wiener-filtered absorption map from the observed \lya\ absorption features \citep{2001Pichon,2008Caucci,2014LeeObserving}. This absorption field can then be related to the underlying matter density through the fluctuating Gunn-Peterson approximation. This Wiener filtering does not explicitly include information about the physical processes of the system and could, in an extreme case, lead to inferred matter distributions which cannot arise from gravitational evolution. In this work, we implement a different approach,
finding the maximum \textit{a posteriori} initial density field which gives rise to the observed density field, often known as a ``constrained realization.''
This will constrain the transmitted flux\footnote{It is a mild misnomer to refer to the Ly$\alpha$ transmission as a `flux', but in this paper we use both 
terms interchangeably.} field to those which are likely to arise from gravitational evolution, providing a more accurate reconstruction at $z=2.5$. This epoch is particularly amenable to this technique since the observed structures are only mildly non-linear and have not yet undergone shell crossing. Not only will this yield information
on the underlying dark matter density field, but also velocity information allowing us to deconvolve redshift space and real space 
quantities (see \citealt{nusser:1999} for a reconstruction method applied to 1-dimensional quasar \lya\ forest sightlines, and \citealt{2001Pichon} for full 3D convolution). This velocity information can also help inform the astrophysical processes occurring in the region; for example combining the flux information, matter velocity information, and a galaxy catalog will provide insights into galaxy formation environmental dependence. In addition, since we have the $z=2.5$ matter density and velocities, we are able to further evolve our 
field to $z=0$ to infer the late time fate of the observed structures. 

%{\color{purple} At some point we should mention that z=2.5 structure is only mildly non-linear and hasn't undergone shell-crossing...}

Reconstructing the initial density field has additional advantages beyond possible improvements in late time reconstruction. As there is currently no evidence for primordial non-gaussianity \citep{2018arXiv180706209P}, the power-spectrum of the initial density modes should provide a loss-less statistic. The entire family of higher order correlations  (such as three-point functions, density peak counts, voids, topological measures, etc.) arise due to gravitational evolution of a density field described by a single power spectra. In the case of galaxy large scale structure surveys, there has already been work towards performing this optimal reconstruction \citep{seljak2017towards}. As \lya\ tomography 
builds up toward cosmological volumes, it would be worth exploring the application of the aforementioned techniques.

In this paper we apply initial density reconstruction to mock observations of IGM Tomography using the Tomographic Absorption Reconstruction and Density inference Scheme (TARDIS). We overview the formalism in Section \ref{sec:method}, describing the optimization scheme, forward model used, and measures of the cosmic web. In Section \ref{sec:mocks}, we describe our mock data-sets which simulate \lya\ tomography observations. In Section \ref{sec:results} we describe our results, and finally discuss next steps in Section \ref{sec:conclusion}.

%It approximately traces the underlying dark matter (DM) (\citep{BiGe}) over-density on large ($\gtrsim 1\,\mpc$) scales making it a probe of large scale structure (LSS) at redshifts beyond those traced by galaxy surveys. With a dense 2 dimensional grid of background faint quasi-stellar objects (QSOs) and bright Lyman-break galaxies (LBGs), it becomes possible to interpolate across 1D sight-lines to `tomographically' reconstruct the 3D \lya\ absorption field which is a biased tracer of the underlying matter field. (\citep{caucci}) 

%The late time fate of these structures is of significant cosmological and astrophysical interest. As baryons follow the dark matter distribution, a detailed understanding of the formation history of the structures would enable detailed hydrodynamic simulations and, in turn, understanding of the observed astrophysical properties. For example, gas shock-heats as it accretes onto protoclusters, and could potentially create cores of reduced Ly$\alpha$ absorption.

%Other dynamic models that have been implemented, such as modified Zeldovich approximation (\citep{TZ}), and augmented Lagrangian perturbation theory (\citep{KH}), seem to only work well in the quasi-nonlinear regime and cannot be accurately extended to the non-linear regime ($k \sim 3.4 \mpc$ at $z=0$). \lya\ Tomography lies in an interesting regime where the most structure is quasi-linear and the ability to reconstruct the initial density field, and late time fate, of observed structures should be more straightforward.

\section{Methodology}
\label{sec:method}

In order to implement our scheme to go fom observed data to the systems initial conditions we need (a) a dynamic forward model (FastPM), (b) an absorption model (FGPA), (c) mapping from field to data-space (flux skewers), (d) a noise model. In this section we describe each component of our model.

\subsection{Modeling}

Here we summarize the optimization technique and standardize notation. For a more complete description, see \citet{seljak1998cosmography,2009simon,seljak2017towards,2018Horowitz}.

We measure $N$ skewers of flux assuming perfect identification of the continuum spectra each of length $L$, and stack those into a full data vector, $\bi{d}$, of total dimension $N \times L$.  This data vector will depend on the initial conditions we wish to estimate at a certain resolution $M$, $\bi{s}$, the Lyman-$\alpha$ absorption model, and a noise term, $\bi{n}$, which we choose to have the same dimension as the data i.e.
\begin{equation}
\bi{d}=\bi{R}(\bi{s}) + \bi{n}
\label{eq:ddecomp},
\end{equation}
where the $\bi{R}: M^3 \rightarrow N\times L$ is the (nonlinear) response operator composed of a forward operator and a skewer-selector function. The Gaussian information is contained in co-variance matrices, $\bi{S}=\langle \bi{s}\bi{s}^\dagger \rangle$, and $\bi{N}=\langle \bi{n}\bi{n}^\dagger \rangle$, for the estimated signal and noise components, which are assumed to be uncorrelated with each other, i.e. $ \langle \bi{n}\bi{(R(s))}^\dagger \rangle = 0$. 
%The correlation matrix of the data is therefore,
%{\color{red}This section seems to move back and forth between a non-linear $R(s)$ and a linear version, $R(s)$.  The below is not true for non-linear $R(s)$ for example:}
%\begin{eqnarray}
%\langle \bi{d}\bi{d}^\dagger \rangle \equiv C &=& \langle (\bi{R}\bi{s}+ \bi{n})(\bi{R}\bi{s}+ \bi{n})^\dagger \rangle \nonumber \\
%&=& \langle (\bi{R}\bi{s} (\bi{R}\bi{s})^\dagger + + \bi{n} \bi{n}^\dagger + \text{Cross Terms} \rangle \nonumber \\
%&=& \bi{R}\bi{S}\bi{R}^\dagger  + \bi{N}.
%\end{eqnarray}
In this work we are interested in maximizing the likelihood of some underlying signal given the data. The generic likelihood function can be written as
\begin{eqnarray}&L(\bi{s}|\bi{d}) = (2 \pi)^{-(N+M)/2} det(\bi{SN})^{-1/2} \times \nonumber \\
& \exp{\left[-\frac{1}{2}\bi{s}^\dagger \bi{S}^{-1} \bi{s} + (\bi{d}-\bi{R(s)})^\dag \bi{N}^{-1}(\bi{d}-\bi{R(s)}) \right]},
\label{eq_likeb}
\end{eqnarray}
where we assume calculate the signal covariance $\bi{S}$ around some fiducial powerspectra. The exponential in this likelihood can be interpreted as a the sum of a prior term ($\bi{s}^\dagger \bi{S}^{-1} \bi{s}$) and a data-dependent term ($(\bi{d}-\bi{R(s)})^\dag \bi{N}^{-1}(\bi{d}-\bi{R(s)})$), with the prefactor as a normalization term. Note that the minimum variance solution for the signal field can be found by minimizing,
\begin{equation}
\chi^2 = \bi{s}^\dagger \bi{S}^{-1} \bi{s} + (\bi{d}-\bi{R(s)})^\dag \bi{N}^{-1}(\bi{d}-\bi{R(s)}),
\label{eq_chi}
\end{equation}
with respect to $\bi{s}$. Working in quadratic order around some fixed $\bi{s_m}$ we have
\begin{equation}
\chi^2 = \chi_0^2 + 2 \bi{g}(\bi{s}-\bi{s_m})+ (\bi{s}-\bi{s_m})\bi{D}(\bi{s}-\bi{s_m}),
\label{eq_chib}
\end{equation}
with gradient function
\begin{equation}
\bi{g}=\frac{1}{2}\frac{\partial \chi^2}{\partial \bi{s}}=\bi{S}^{-1}\bi{s_m}-\bi{R'}^\dag\bi{(s_m)}\bi{N}^{-1}(\bi{d}-\bi{R(s_m)}),
\label{eq_cost}
\end{equation}
and curvature term\footnote{Note we drop the $\bi{R''}$ term as it fluctuates with mean zero and doesn't appreciably affect the optimization.} 
\begin{equation}
\bi{D}=\frac{1}{2}\frac{\partial^2 \chi^2}{\partial \bi{s}\partial \bi{s}}=\bi{S}^{-1}+\bi{R'}^\dag\bi{N}^{-1}\bi{R'}. %+ \bi{R''}\bi{N}^{-1}(\bi{d}-\bi{R}\bi{s_m}).
\label{eq_curv}
\end{equation}
Calculation of the derivative term $\bi{R'}$ requires calculation with respect to every initial mode. We use an automated differentiation framework in
Appendix B of \citet{2018fengseljakzaldarriaga} 
o calculate Jacobian products of our evolution operator without running additional simulations. This avoids running additional involved simulations with respect to every mode, which would be prohibitively costly.

\subsection{Optimization }

%{\color{red}This section is a huge, disconnected jump from the previous section and doesn't seem to have much of a point -- it's largely name-dropping at present.  I think most of the content is here, you just need to organize it better.}

As each iteration of the chain requires running a PM simulation, it is important to minimize computational time. While others have used Hamiltonian Markov Chain Monte Carlo (HMC) algorithms to find fast reconstructions for galaxy surveys \citet[see][]{HMC,ECULID1,Eculid3}, in this work we are instead finding the most likely map reconstruction. We therefore use a Limited-memory Broyden – Fletcher – Goldfarb – Shanno (LBFGS) algorithm \citep{NumRec}, a general technique for solving nonlinear optimization problems. Rather than sampling over the entire parameter space, LBFGS takes a quasi-Newtonian approach, i.e. it is similar to the standard Newton-Ralphson method but rather than calculating the inverse of the entire Hessian (a very large matrix for a density field on the scales of interest) it iteratively updates a pseudo-Hessian as the function is being optimized.

Quasi-Newtonian methods, like L-BFGS, are only guaranteed to find extrema for convex optimization problems. For the case of large scale structure, it was demonstrated that the posterior surface is multimodal at the smallest scales but not modes probed by next generation large scale structure surveys \citep{2018fengseljakzaldarriaga}. This optimization technique was previously implemented for the case of cosmological shear measurements and CMB reconstruction, finding fast numerical conversion even in very high dimensional parameter space \citep{2018Horowitz}, as well as in dark-matter-only models \citep{seljak2017towards,2018fengseljakzaldarriaga}.

Our implementation is based on the \texttt{vmad} framework,\footnote{https://github.com/rainwoodman/vmad} an extension of the \texttt{abopt} framework used to perform similar reconstructions from late time galaxy fields \citep{2018Chirag}. This framework allows very fast reconstruction convergence; for cases studied in this work each reconstruction took approximately 5 CPU-hours. 

\subsection{Response Function and Forward Model}

\begin{comment}
\begin{figure}[t]
\begin{center}
\begin{overpic}[width=0.32\textwidth]{./figs_fastpm/forward/ic.png}
\put(25,-10){\textsf{\scriptsize (a) Initial Density Field }}
\end{overpic}
\begin{overpic}[width=0.32\textwidth]{./figs_fastpm/forward/rho.png}
\put(14,-10){\textsf{\scriptsize (b) Evolved Density Field, Real Space}}\end{overpic}
\vspace{1em}
\begin{overpic}[width=0.32\textwidth]{./figs_fastpm/forward/rho_z.png}
\put(25,-10){\textsf{\scriptsize (c) Evolved Density Field, Redshift Space}}
\end{overpic}
\end{center}
\vspace{-0.4cm}

\begin{center}
\begin{overpic}[width=0.495\textwidth]{./figs_fastpm/forward/flux.png}
\put(30,-10){\textsf{\scriptsize (d) Flux Field in Redshift Space}}
\end{overpic}
\vspace{1em}
\begin{overpic}[clip,trim={0cm 0cm 0 0cm},width=0.495\textwidth]{./figs_fastpm/forward/vel.png}
\put(20,-10){\textsf{\scriptsize (e) Velocity along LOS}}
\end{overpic}
\end{center}

\caption{\label{fig:512CMB}
Slice of forward model (steps 1-5). [I'll make a slightly prettier picture...]}
\end{figure}
\end{comment}

Optimization over the initial density skewers requires defining a differential forward model which will allow us to define a $\chi^2$ problem as in Eq. \ref{eq_chi} and gradient function as in Eq. \ref{eq_cost}.

\subsubsection{Forward Evolution}

Following the work of \citet{2018fengseljakzaldarriaga} we first use Lagrangian Perturbation Theory (LPT) to evolve the initial conditions while the field is still almost entirely linear. We do this until $z=100.0$, at which point we then use 5 steps of FastPM \citep{fastPM}\footnote{https://github.com/rainwoodman/fastpm} to evolve until redshift $z=2.5$. 

There are fundamental limitations due to using a particle mesh framework with limited time steps, constraints imposed by the speed requirements for optimization. As discussed in \citet{fastPM} and \citet{2018Dai}, halos are not fully virialized when using these methods. This will not affect our ability to reconstruct structure on $>1$ \mpc scales
relevant for current and upcoming surveys. Similarly, we use a particle resolution of $128^3$ for our reconstructions to allow fast optimization.

We use the $z=2.5$ particle positions to generate a density field and infer the hydrogen Ly$\alpha$ optical depth using the Fluctuating Gunn Peterson Approximation (FGPA), with $T = T_0 (\rho/\bar{\rho})^{(\gamma - 1)}$ with slope $\gamma = 1.6$ \citep{lee:2015PDF}. Note that we calculate the optical depth first, which is then redshift-space distorted using the inferred velocity field. Then we compute the flux $F=\exp\left(-\tau\right)$ and select lines of sight matching the positions of the mock observations. The skewers are then smoothed with a $\sigma=1.0$ Mpc/h Gaussian filter to imitate spectrographic smoothing; this is a conservative estimate for upcoming surveys.

\begin{figure*}
    \centering
    \includegraphics[width=0.95\textwidth]{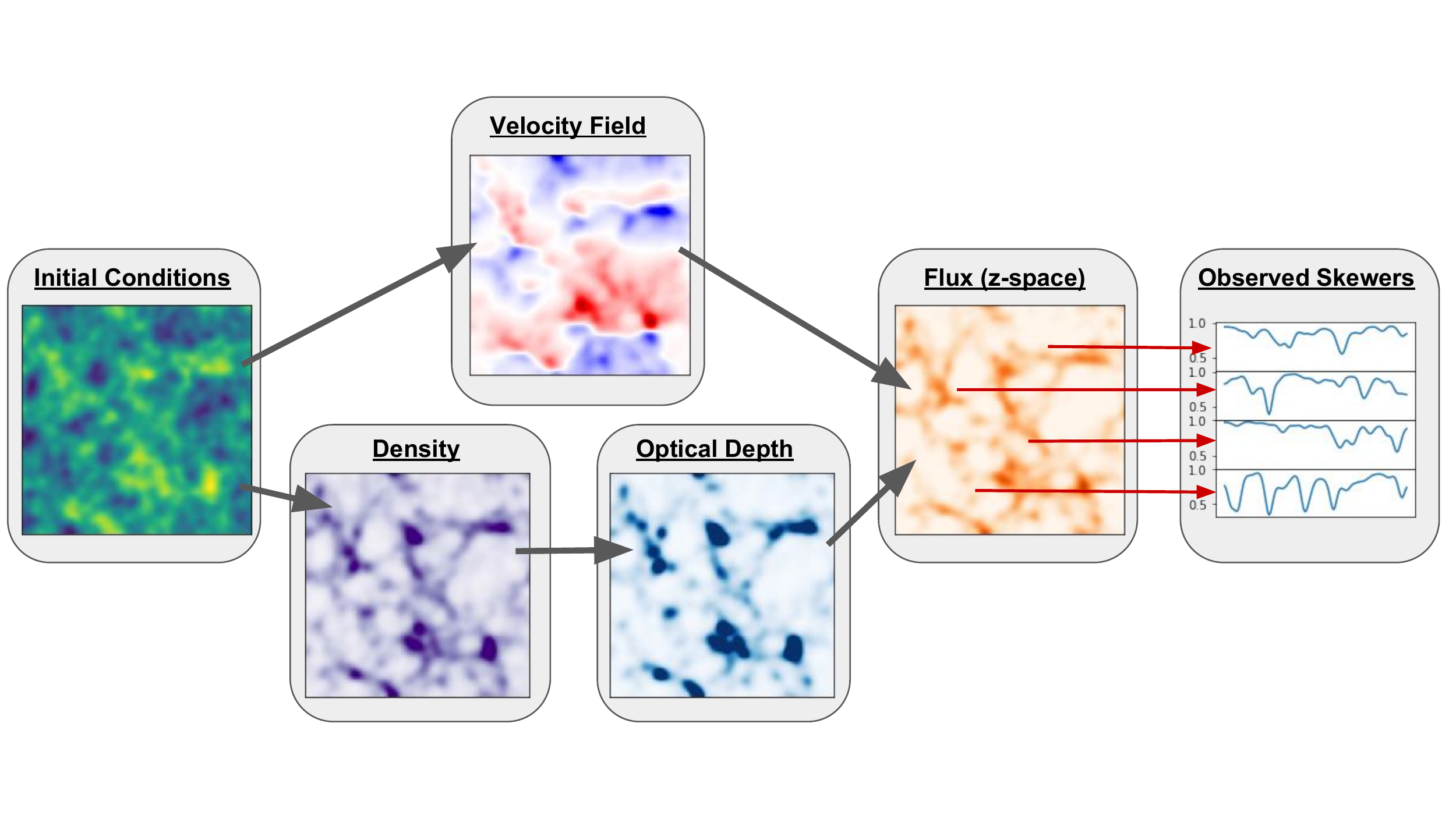}
\caption{\label{fig:overview}
Schematic illustration of our forward model (see Sec~\ref{subsubsec_forwardmodel}). The underlying field we are optimizing for is the initial matter density field (left). 
The output of our forward model are the \lya\ flux skewers probing the observational volume at the same positions as the data.}
\end{figure*}

\subsubsection{Overview of Forward Model}
\label{subsubsec_forwardmodel}

\begin{enumerate}
\item Initialize a Gaussian random field (the signal field).
\item Evolve field forward to $z = 2.5$ with FastPM.
\item Use FGPA to calculate a real space Ly$\alpha$ optical depth.
\item Use the line of sight velocity field to shift the Ly$\alpha$ optical depth to redshift space.
\item Exponentiate the redshift space optical depth field to get the transmitted flux field.
\item Select skewer sightlines from redshift space flux field.
\item Convolve skewers with Gaussian spectrograph smoothing. 
\end{enumerate}

\section{Mock Datasets}

\label{sec:mocks}

\begin{deluxetable*}{c c c c c c l}
    \tablecolumns{7}
    \tablecaption{\label{tab:table1} Mock Data Sets for Reconstructions}
   \tablehead{ \multirow{2}{*}{Name} & \colhead{N-body} & \colhead{LOS Separation} & \colhead{LOS Density} & \colhead{S/N$_\textrm{min}$} 
   & \colhead{S/N$_\textrm{max}$} & \multirow{2}{*}{Description} \vspace{-0.5em} \\ 
    & \colhead{Code} & \colhead{($\mpc$)}  & \colhead{(deg$^{-2}$)} & \colhead{ (\r{A}$^{-1}$)} & \colhead{(\r{A}$^{-1}$)} & }
   	%\colhead{\multirow{2}{*}{Name & }} & \colhead{ \multirow{2}{*}{N-Body & Code}} &
	%\colhead{\multirow{2}{*}{LOS Separation & ($\mpc$)}} & \colhead{\multirow{2}{*}{Sightline Density & (deg$^{-2}$)} }  &
	%\colhead{\multirow{2}{*}{S/N$_\textrm{min} $ & ($\AA^{-1}$)} }  &  \colhead{\multirow{2}{*}{S/N$_\textrm{max} $  & ($\AA^{-1}$)}} &
	%\colhead{\multirow{2}{*}{Volume & ($h^{-3}\,\textrm{Mpc}^3$)} }   }
	\startdata
        \texttt{T-TomoDESI} & TreePM & 3.7 & 363 & 1.4 & 4.0 & Dedicated survey with DESI spectrograph (4m)\\
        \texttt{T-CLA/PFS} & TreePM & 2.4 & 863 & 1.4 & 10.0 & Survey with 8-10m-class telescopes \\
        \texttt{T-30+T} & TreePM & 1.0 & 4970 & 2.8 & 10.0 & Survey with 30m-class telescopes \\
        \texttt{F-CLA/PFS} & FastPM & 2.4 & 863 & 1.4 & 10.0 & Same as  \texttt{T-CLA/PFS}, but using FASTPM \\
    \enddata
\end{deluxetable*}

While the FastPM code provides a rapid convergence towards the underlying density field within the TARDIS framework, 
to rigorously test our reconstruction we apply the formalism to mock data generated from well-characterized large-volume, high-resolution N-body simulations.
We therefore use a simulation volume run with TreePM \citep{2002White,2010White}, which has been used for other work on Lyman-$\alpha$ forest tomography \citep{2015StarkProtocluster,Stark2015,2018Krolewski} This simulation uses $2560^3$ particles in a box with 256 \mpc along each dimension, with cosmological parameters $\Omega_m = 0.31$, $\Omega_b h^2 = 0.022$, $h = 0.677$, $n_s = 0.9611$, and $\sigma_8=0.83$. The initial conditions are generated using second order Lagrangian Perturbation Theory to $z_{ic}=150$ and then further evolved using the TreePM code. The output was taken at $z=2.5$ and $z=0$ for comparison, and a $z=2.5$ Lyman-$\alpha$ absorption field was generated using the FGPA with $T_0 = 2.0 \times 10^4$ and $\gamma=1.6$. 

We generated mock skewers from $(64\,\mpc)^3$ subvolumes of the TreePM simulation
with different survey parameters to mimic various ongoing and upcoming 
IGM tomography surveys --- these are summarized in Table~\ref{tab:table1}. 
The most important survey parameter is the mean sightline separation, or equivalently areal
density of background sources on the sky. This is typically set by the overall sensitivity of the telescope/instrument
combination and desired integration time, but in this work we simply quote the sightline separation and
minimal S/N for each survey; we refer the 
reader to \citet{Lee2014Theory} for a more detailed discussion with respect to observational strategy.
The CLAMATO survey \citep{Lee2017}, which is currently ongoing with the Keck-I telescope, achieves a mean separation of $2.4\,\mpc$ between 
sightlines (albeit over a small footprint of 0.16 deg$^2$ at present). 
An IGM tomography program is currently being planned for the upcoming Prime Focus Spectrograph \citep{sugai:2015}, which should achieve comparable spatial sampling
as CLAMATO but over a much larger area ($\sim 15$ deg$^2$). 
Further into the 2020s, thirty-meter class telescopes such as TMT, ELT, and GMT will allow much greater sightline densities by observing fainter background sources.
While the exact parameters of future IGM tomography surveys on thirty-meter telescopes will depend on instruments that are largely still under early development, 
for now we assume a $1\,\mpc$ sightline separation. 
We also study a hypothetical dedicated
 IGM tomography program carried out with the DESI spectrograph, which is currently 
being installed on the 4m Mayall telescope \citep{desi}. Note that this is \textit{not} 
the quasar \lya\ forest survey currently being planned as part of the DESI cosmology program, which at
only $\sim 50-60\,\mathrm{deg}^{-2}$ it is far too sparse for cosmic web analysis.
While the DESI instrument offers 5000 fibers over a 7.5 deg$^2$ field-of-view, we assume that 10\% of the fibers will be dedicated to sky subtraction
and a $1.7\,\times$ overhead factor in background sources targeted to maintain the specified sightline density over a finite redshift range of $\delta z=0.3$ \citep{Lee2014Theory}. This implies a mean
sightline separation of $3.7\,\mpc$ for a dedicated DESI tomography program.

For pixel noise, we assume Gaussian random noise which varies among different skewers but is constant along 
each skewer.
To simulate a realistic distribution of skewer S/N, we follow the
prescriptions in \citet{Stark2015} and \citet{2018Krolewski} and draw the individual skewers' S/N from a power-law distribution with minimum value S/N$_\textrm{min}$ (i.e.\ $\rm{d}n_\textrm{ los}/\rm{d}\textrm{S/N} \propto \textrm{S/N}^{\alpha}$) and spectral amplitude $\alpha=2.7$. The S/N$_\textrm{min}$ is the same for both the DESI and CLAMATO/PFS mocks since it reflects the
actual minimal S/N in the real CLAMATO data, but for 30m-class telescopes \citet{Lee2014Theory} found that
the S/N needs to be increased as the tomographic reconstruction is no longer limited by the shot-noise from finite skewer sampling.
To be conservative, we also impose a maximal S/N for all mock datasets \citep{Lee2017}, as specified in 
Table~\ref{tab:table1}.

\begin{figure*}
  \centering  \includegraphics[trim=0cm 0cm 0cm 0cm,width=0.98\textwidth]{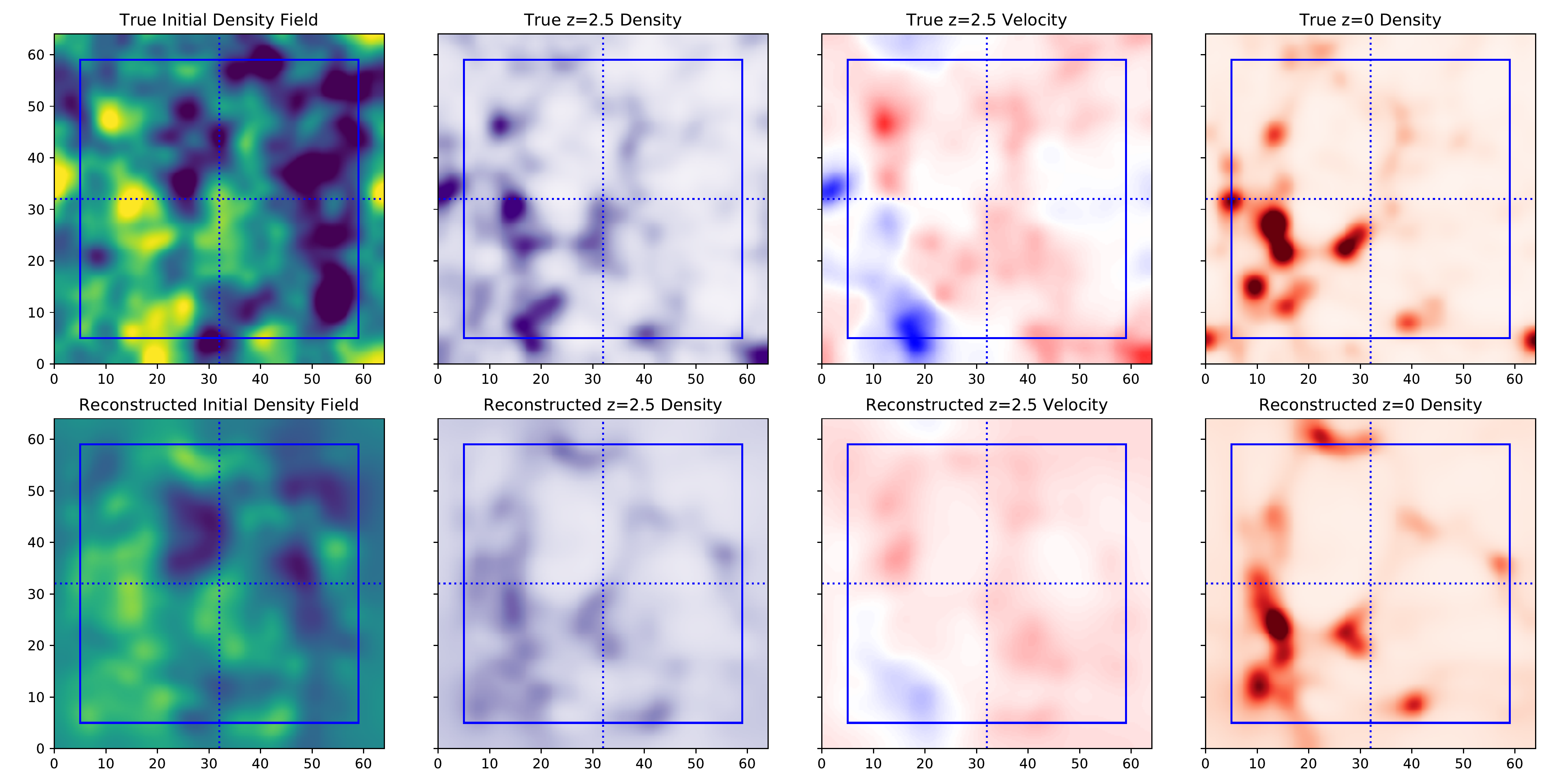}
    \caption{Reconstruction of various recovered quantities for the \texttt{F-CLA/PFS} mock dataset, smoothed at 2 \mpc, are shown on the bottom row. The true corresponding fields from the
    FastPM simulation are shown on the top. In all panels we project along a 5 \mpc slice. The region outside the solid blue box is masked in our analysis, while the dotted
    lines are merely to guide the eye. We find that the large scale features are qualitatively captured well in the reconstructions.
    \label{fig:8panel}}
\end{figure*}

In addition to the random pixel noise, we add  continuum error to account for the difficulty in identifying the intrinsic quasar or galaxy continuum. The ability to estimate the continuum is dependent on the S/N of the skewers, and we apply the fitted continuum error distribution of \citet{2018Krolewski} to our mock skewers. In particular, we take our observed flux to be 
\begin{equation}
    F_\textrm{obs} = \frac{F_\textrm{sim}}{1+\delta_c},
\end{equation}
where $\delta_c$ is taken from an underlying Gaussian distribution
with width $\sigma_c$ depending on S/N along each skewer as
\begin{equation}
 \sigma_c = \frac{0.205}{\rm{S/N}} + 0.015
\end{equation}
where the constants are fitted from data from the CLAMATO field. While we add continuum
errors to our mock spectra, we do not
directly model continuum error in TARDIS. This could be included as an off-diagonal term in the covariance matrix in future work.

In addition to the TreePM run, we have also generated mock skewers from FastPM using the exact same technique and parameters as in our forward model. This serves to isolate effects caused by known limitations of FastPM to resolve small scale halo properties, as well as provide a tool for rapid consistency checks. These are applied towards the
discussions regarding the code convergence in Appendix \ref{app:converge}, and the method's sensitivity to astrophysical assumptions (Appendix \ref{app:sens}).

\section{Results}
\label{sec:results}

We apply the TARDIS method, described in \S\ref{sec:method}, to the mock data set generated as described in \S\ref{sec:mocks}. 
Broadly, we are interested in how well we reconstruct cosmic structures both at the observed redshift ($z=2.5$) and the late time ($z=0$) fate of those structures. 
TARDIS solves for the initial density fluctuations within the volume, which one can then use to initialize a simulation using any cosmological N-body or hydrodynamical
code to study the cosmic evolution of the large-scale structure realization. 
For convenience, however, in this paper we continue to use FastPM to study the gravitational evolution of the TARDIS realizations at both $z=2.5$ and $z=0$. 
The $z=2.5$ field simply the best-fit TARDIS solution, whereas to get to $z=0$ we evolve FastPM by another five steps.
We then compare the resulting fields with the `truth' from the fiducial TreePM simulation volume.

Examples of reconstructed fields for initial density, $z=2.5$ matter density and \lya\ flux, line of sight velocity and $z=0$ matter density for \texttt{T-CLA/PFS} are shown in Fig \ref{fig:8panel}. 
In comparison with the `true' fields, there is a strikingly good recovery of the overall filamentary
backbone of the $z=2.5$ matter density field as well as the overall distribution of the velocity field. However, the TARDIS reconstruction appears to underestimate the overall
amplitude of the density field, with less prominent density peaks in both the initial conditions
and $z=2.5$ matter density. As expected, the underestimated matter power propagates through
to the evolved density field at $z=0$, where the density peaks in the reconstruction are much 
less prominent than the true underlying density.

\begin{figure}\centering
  \includegraphics[trim=1cm 0cm 0cm 0cm,width=0.40\textwidth]{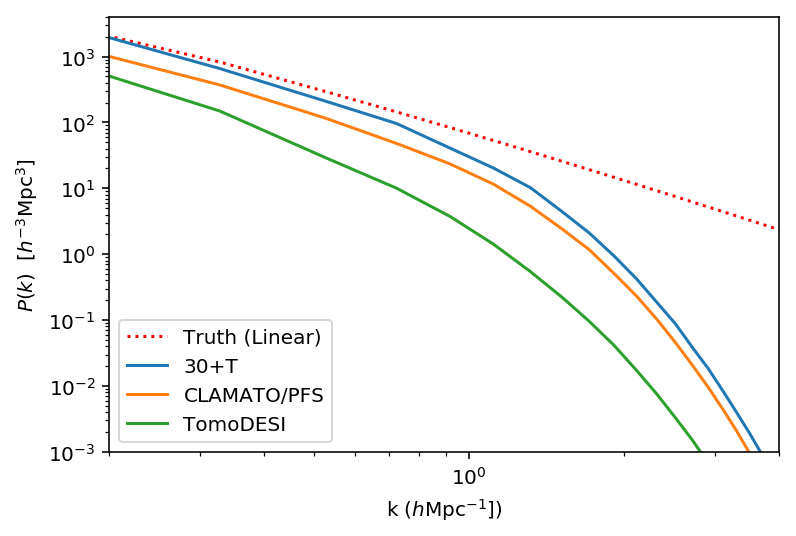}
    \includegraphics[trim=1cm 0cm 0cm 0cm,width=0.40\textwidth]{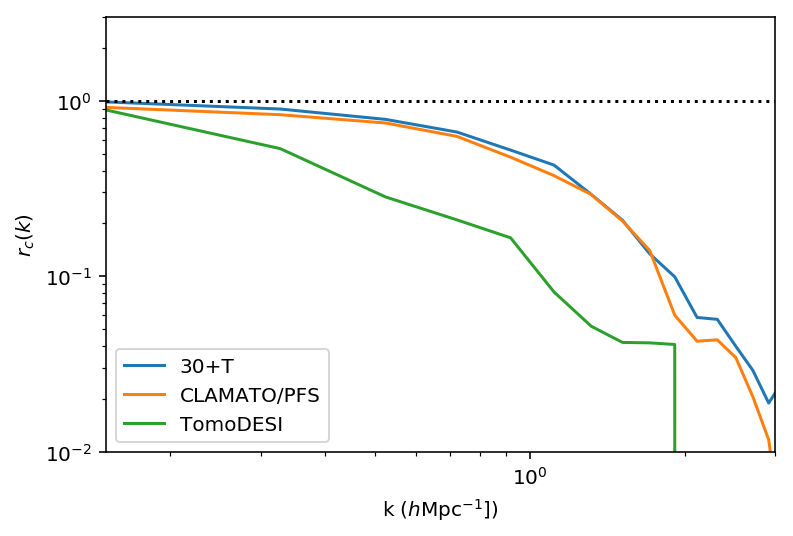}
    \caption{\emph{Top:} Power spectra of the reconstructed initial conditions for various experimental configurations, with the true initial conditions shown for comparison. \emph{Bottom:} Cross correlation coefficient, $P_{RT}/\sqrt{P_{RR}P_{TT}}$, where $P_{RT}$ is the crosspower between the true field and reconstructed field, $P_{RR}$ is the reconstructed power spectrum and $P_{TT}$ is the true power spectrum. As the number of sight-lines and spectral noise improve, power spectra reconstruction improves; however there remains a residual noise bias for realistic experiments.}
    \label{fig_ps}
   % \afterpage{\FloatBarrier}
\end{figure}

The underestimated matter amplitude appears to be a result of the reconstruction method, and can
be seen when we compare the reconstructed initial fluctuation power spectrum with that used to generate the
`true' TreePM simulation volume (Figure~\ref{fig_ps}). There is a shortfall in the recovered power
in all the mock reconstructions, especially on scales below the mean sightline density
of the mock data, but also on larger scales. This gets worse with reduced sightline density of the
\texttt{T-TomoDESI} reconstruction, while conversely the improved sightline sampling of the \texttt{T-30+T} mock allows a better
job of recovering the true power spectrum, although there is still a shortfall at all scales.
This is possibly due to the fact that the \lya\ forest absorption
blends and saturates in matter overdensities. In particular, at a fixed noise level, \lya\ forest features have higher density resolution at lower absorption levels than at higher absorption levels due to the exponential FGPA mapping. For example, the optimization algorithm can distinguish between a $1 \sigma$ and $2 \sigma$ overdensity at higher significance than a $10 \sigma$ and $11 \sigma$ overdensity at a given flux noise-level. While it might be possible to correct for this reduced power in the initial density fluctuations,
this is a non-trivial process which we defer to an upcoming paper that will focus on modeling galaxy 
protoclusters within the TARDIS framework. It is also possible to adjust for this non-linear noise bias at the power-spectra level within a response formalism \citep{seljak2017towards,2018Horowitz}.

Nevertheless, TARDIS appears to do a reasonable job in recovering the moderate-density cosmic web
as seen in Figure~\ref{fig:8panel}. 
We thus focus on the large-scale cosmic web, and compare the performance of the TARDIS across cosmic
time.

\subsection{Classification of the Cosmic Web}
\label{subsec_classificationcosmicweb}

For quantitative comparison of the large-scale structure recovery in TARDIS, we use the deformation tensor cosmic web classification of \citet{2017Krolewski} and described in \citet{2016LeeWhite}, which was inspired by \citet{Bond:1993,2007Hahn,forero-romero:2009}. While there exist other cosmic web classification algorithms \citep[see summary in ][]{2014Cautun}, the deformation tensor approach has a strong physical interpretation within the Zel'dovich approximation \citep{1970Zeldovich} and allows easy comparison to previous work in the context of Lyman-$\alpha$ forest tomography. However, in contrast to \citet{2016LeeWhite} and \citet{2017Krolewski}, who measured the eigenvalues and eigenvectors of Wiener-filtered maps of the \lya\ transmitted flux, in this work we directly measure the eigenvalues and eigenvectors of the dark matter fields reconstructed with TARDIS, which have been first smoothed with a $R=2\,\mpc$ Gaussian kernel.

 %\newpage

\begin{figure*}[t]
  \begin{center} 
  \includegraphics[trim=0cm 0cm 0cm 0cm,width=0.90\textwidth]{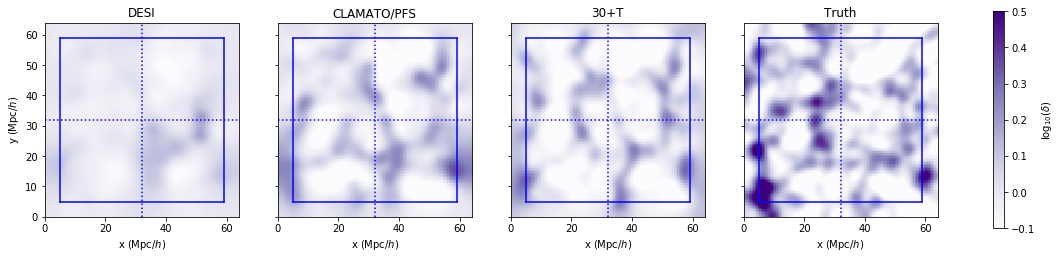}
  \end{center}
  \hspace{26.5pt}
  \includegraphics[trim=0cm 0cm 0cm 0cm,width=0.80\textwidth]{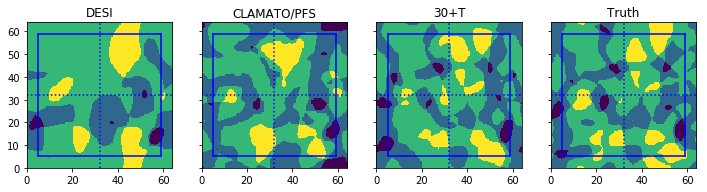}
    \caption{Comparison of the $z=2.5$ reconstructed cosmic structures as classified by their eigenvalues, from \texttt{T-TomoDESI}, \texttt{T-CLA/PFS}, and \texttt{T-30+T}, vs. the true $z=2.5$ density field for an \emph{xy}-slice. Fields have been smoothed by a $R=2 \mpc$ Gaussian kernel. \emph{Top}: matter density. \emph{Bottom}: classification of cosmic structure. Dark blue indicates node, light blue indicates filament, green indicates sheet, and yellow indicates void. The region outside the solid blue box is masked in our analysis, while the dotted
    lines are to guide the eye.
    We find our classification captures the visual appearance of the cosmic web well and that the recovered structure improves as number of sight-lines increases and noise decreases. 
    \label{fig_config}}
  \centering 
  \includegraphics[trim=1cm 0cm 0cm 0cm,width=0.7\textwidth]{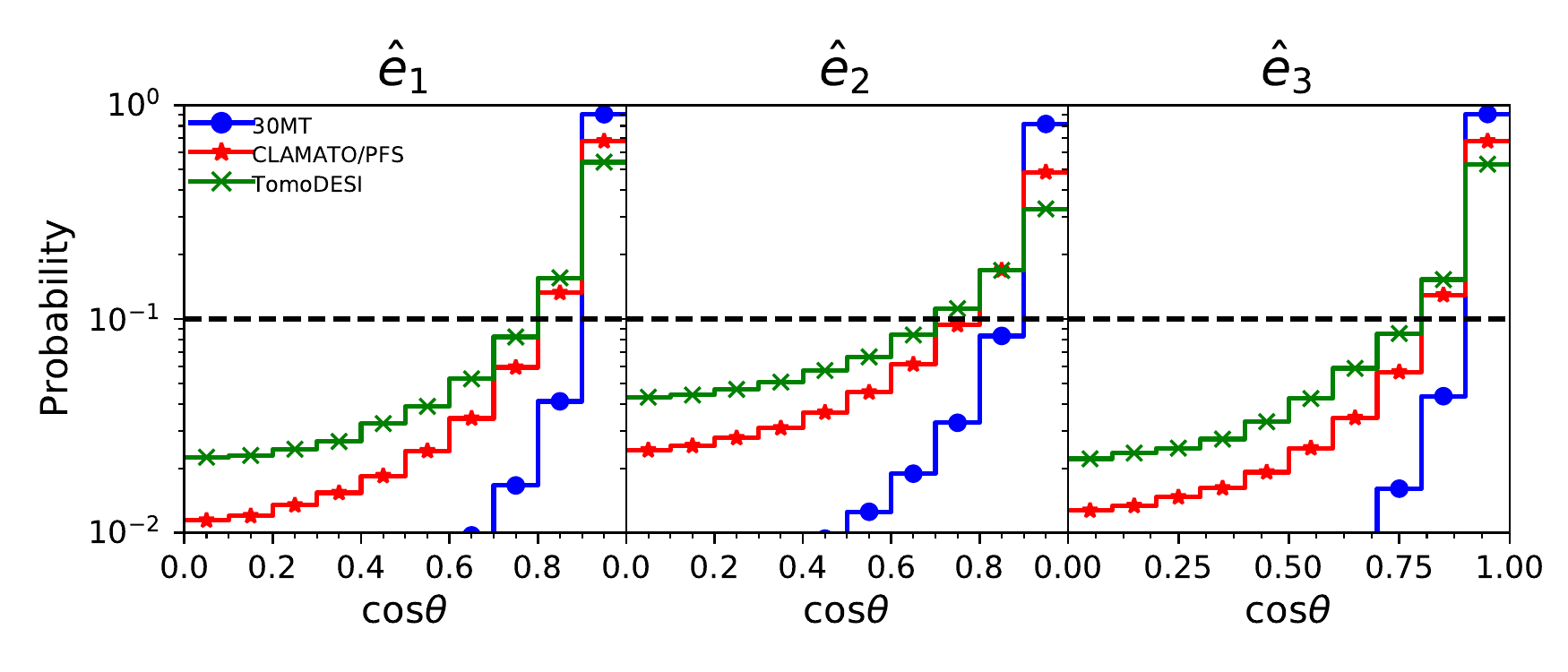}
    \caption{\label{fig_cosmicweb} PDF showing the dot product of the eigenvectors from cosmic web reconstruction vs. the true cosmic web for various experimental configurations. $\cos\theta = 1.0$ indicates the cosmic web structures are oriented the same way, while $\cos\theta = 0.0$ indicates perpendicular alignment. Horizontal dashed line indicates the expected
    distribution for randomly aligned structure. In \texttt{T-30+T} the recovery of the cosmic web structure is near perfect, with only very slight misalignments on average.
    }
\end{figure*}

The eigenvectors and eigenvalues of the deformation tensor relate directly to the flow of matter around that point in space; matter collapses along the axis of the eigenvector when the associated eigenvalue is positive, and expands when it is negative. Points with three eigenvalues above some nonzero threshold value $\lambda_{th}$ \citep[as in][]{forero-romero:2009} are nodes (roughly corresponding to (proto)clusters), two values above $\lambda_{\rm th}$ are filaments, one value above $\lambda_{\rm th}$ are sheets, and zero values above $\lambda_{\rm th}$ are voids. The deformation tensor, $D_{ij}$, is defined as the Hessian of the gravitational potential, $\Phi$, i.e.

\begin{equation}
    D_{ij} = \frac{\partial^2 \Phi}{\partial x_{i} \partial x_{j}},
    \label{eq:diften}
\end{equation}
or equivalently in Fourier space in terms of the density field, $\delta_k$, as
\begin{equation}
    \tilde{D}_{ij} = \frac{k_i k_j}{k^2}\delta_k.
    \label{eq:diften_k}
\end{equation}
This tensor is then diagonalized to obtain the eigenvalues $\hat{e}_1$, $\hat{e}_2$, and $\hat{e}_3$ at each point on our spatial grid, ordered such that their corresponding eigenvalues are $\lambda_1>\lambda_2>\lambda_3$ (i.e. to demand that collapse first occurs along $\hat{e}_1$). Note that one could use the velocity field from the reconstruction itself to determine the flow at each point
\citep[e.g.][]{2013Libeskind,2016Pahwa} instead of relying on the Zel'dovich approximation used in the classification here. We use the deformation tensor in order to stay consistent with past IGM tomography work \citep{2016LeeWhite,2017Krolewski}.
Cosmic web directions for our reconstructed field are thus defined by the eigenvectors with associated eigenvalues used to classify the cosmic web.

We follow \citet{2017Krolewski} and \citet{2016LeeWhite} and define our threshhold value $\lambda_{th}$ for each simulated field such that the voids occupy 21\% of the total volume at $z=2.5$ and 27\% at $z=0$ \citep[inspired by the redshift evolution in][]{2014Cautun}. The void fraction is somewhat arbitrary in the analysis, as long as it is consistent between the mock reconstructions and true density field used for comparison.

\begin{figure*}
\centering
  \includegraphics[trim=1cm 0cm 0cm 0cm,width=0.80\textwidth]{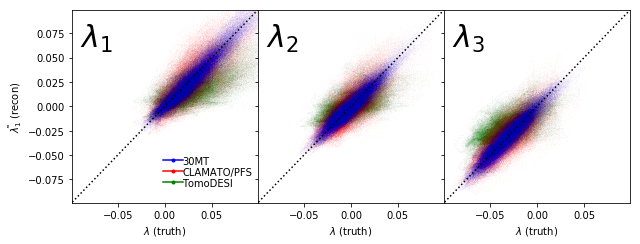}
    \caption{The point by point distribution of the eigenvalues inferred from the deformation tensor, smoothed by 2 \mpc. The magnitude of each eigenvalue indicates the magnitude of compression along the associated eigenvector. As sight-lines increase and noise decreases not only is there less scatter in the eigenvalues, but also less overall bias.
    \label{fig_eigenvalues}}

  \centering 
  \includegraphics[trim=1cm 0cm 0cm 0cm,width=0.8\textwidth]{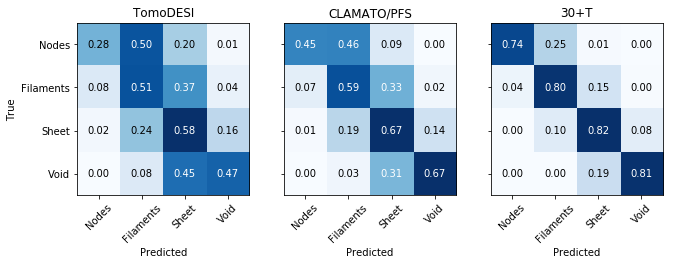}
    \caption{\label{fig_confusion_z=2.5} Confusion matrix for cosmic structures at $z=2.5$ in real space showing with the reconstructed fraction printed over each cell. For \texttt{T-30+T}, we correctly identify approximately 80\% of the volume.} 
\end{figure*}

\begin{deluxetable*}{@{\extracolsep{6pt}} l  c c c   c c c c @{}}
\tablecolumns{8}
\tablecaption{\label{tab:table1} Cosmic Web Recovery at $z=2.5$ (Eulerian Comparison) }
\tablehead{ \multirow{2}{*}{Mock Data}
   & \multicolumn{3}{c}{Pearson Coefficients} & \multicolumn{4}{c}{Volume Overlap (\%)} \\ \cline{2-4} \cline{5-8}
    \colhead{}  & \colhead{$\lambda_1$} &  \colhead{$\lambda_2$} & \colhead{$\lambda_3$} & 
      \colhead{Node} & \colhead{Filament} & \colhead{Sheet} & \colhead{Void} }
\startdata
        \texttt{T-TomoDESI} & 0.62 & 0.58 & 0.66 & 28 & 51 & 58 & 47 \\
        \texttt{T-CLA/PFS}& 0.78 & 0.75 & 0.77 & 45 & 59 & 67 & 67 \\
        \texttt{T-30+T} & 0.94 & 0.94 & 0.95 & 74 & 80 & 82 & 81
    \enddata
  %      \caption{Pearson correlation coefficients and volume overlap fractions for the three mock datasets, 2 \mpc at z=2.5.}
\end{deluxetable*}

\subsection{Matter/Flux Density at $z\sim 2.5$}
\label{subsec_z=2.5}
We compare the recovery of $z=2.5$ \lya\ flux
to previously-standard Wiener filtering techniques.
As we are assuming the Fluctuating Gunn Peterson approximation, this reconstructed flux can be mapped directly to the density field. While past work on Wiener-filtered IGM tomographic maps \citep{Lee2017,2008Caucci} have smoothed the field on $1.4\times$ the mean sightline spacing, for these comparisons we smooth the respectively matter fields
with a $\sigma=2\,\mpc$ Gaussian kernel. The smaller smoothing scale is appropriate for our work because our method should be better able to infer nonlinear and semi-linear structure between sight-lines. For all plots we treat the field in real space (without redshift space distortions) since our optimization is over the initial real space density field.
%\onecolumngrid 

The reconstructed matter density fields from the various mock IGM tomography surveys (summarized in 
Table~\ref{tab:table1}) are shown in the first row of Figure~\ref{fig_cosmicweb}
in comparison with the true density field from the TreePM simulation.
In all cases, they are smoothed with a $R=2\,\mpc$ Gaussian kernel. On large scales, 
the reconstructed density fields are well matched in terms of voids and sheets, 
but CLAMATO/PFS data misses out on some prominent filamentary structures and nodes
as a consequence of the underestimated matter amplitude.
The 30+m telescopes, on the other hand, yield an matter density reconstruction with
excellent fidelity over the entire volume. 

We next calculate the characteristic eigenvalues of the deformation tensor, as described
in \S~\ref{subsec_classificationcosmicweb}, on the smoothed matter density fields.
The scatter of the eigenvalues, relative to the true underlying eigenvalues, is plotted in Figure~\ref{fig_eigenvalues}. This reflects how well recover the amplitude of curvature of 
the matter density field along each cosmic web direction. The distribution of all three eigenvalues is unbiased relatively to the truth, 
albeit with more scatter in the case of the sparser CLAMATO/PFS reconstruction.
We quantify the agreement in terms of Pearson correlation coefficients, showing the scatter from a linear trend in Table~\ref{tab:table1}.
These show a strong correlation between the reconstructed and true eigenvalues, ranging from 
$r=[0.78, 0.75, 0.77]$ in recovering the three eigenvalues $[\lambda_1, \lambda_2, \lambda_3]$ for CLAMATO/PFS, to the excellent
reconstruction of the 30m-class telescopes with correlation coefficients of $r=[0.94, 0.94, 0.95]$.

Next, we classify the each point within the density field as void, sheet, filament, or node depending on how many of the eigenvalues are greater than the threshold value,
$\lambda_i> \lambda_\mathrm{th}$. 
In the true matter density field, we find that $[22, 50, 25, 3]\%$ of the volume is occupied by voids, sheets, filaments, and nodes, respectively --- by construction the reconstructed
matter fields show similar volume occupation fractions to within $\pm 2\%$. 
The volume overlap fraction between cosmic web classifications in the mock data reconstructions compared to the true matter field are listed in Table~\ref{tab:table1} --- these do not include a buffer region of $5$ \mpc near the edge of the volume where we expect to 
be contaminated by  boundary effects.
For the CLAMATO/PFS mock reconstructions, the volume overlap fractions are $\sim 60-62\%$ for the sheets and voids, declining
to $32\%$ for the nodes. It is unsurprising that the nodes are more challenging to recover, since they occupy such a small
fraction (3\%) of the overall density field. These numbers are, on the surface, comparable to those found by \citet{2017Krolewski} 
(their Table~1) for
a similar CLAMATO-like mock data set, but in fact somewhat better since we are probing the matter field directly
on $2\,\mpc$ scales, whereas \citet{2017Krolewski} were evaluating the \lya\ transmission field over coarser ($4\,\mpc$) scales in the 
equivalent case. 
This improvement is due to the fact that the TARDIS incorporates the physics of gravitational evolution into its
reconstructions, in contrast with Wiener-filtering, which only assumes a correlation function.
The 30m-class reconstruction, as expected, fares even better thanks to its finer sightline sampling, with the voids, sheets, 
filaments, and nodes overlapping $[81\%, 82\%, 80\%,  74\%]$ with the true matter density cosmic web.

\begin{figure}
        \begin{center}
    \includegraphics[width=0.34\textwidth]{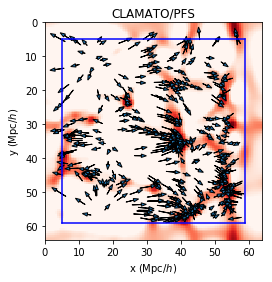}
    \includegraphics[width=0.34\textwidth]{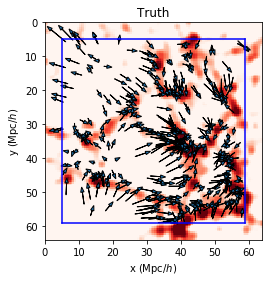}
\end{center}
    \caption{Displacement fields from $z=2.5$ to $z=0$ for random matched particles between the TreePM truth and the reconstructed in the mock observed volume. The underlying $z=0$ density field is also shown. TARDIS is able to well reconstruct the movement and $z=0$ environment of test particles identified at $z=2.5$.
    \label{fig_displacement}}
\end{figure}

To further illustrate the fidelity of the recovery, Figure \ref{fig_confusion_z=2.5} shows the confusion matrix, 
evaluated at all the grid points in our volume, between the true cosmic web from the simulation and our reconstructions, 
finding good agreement. Overall, we find 80\%, 60\%, and 53\% of the total observed volume is properly classified for \texttt{T-30+T}, \texttt{T-CLA/PFS}, and \texttt{T-TomoDESI}, respectively. Allowing mis-classification by a structurally adjacent type (i.e. void to sheet, sheet to void/filament, filament to sheet/node, and node to filament) the agreement goes up to 98\%, 96\%, and 95\% respectively.
We also examine the eigenvector recovery by computing the dot product 
between the eigenvectors recovered from the reconstructions with those at the same 
Cartesian point in the true matter density field\footnote{These values only include structure in the observed region, excising an additional buffer of 2 \mpc near the survey boundary.} --- with a good recovery, the recovered eigenvectors would be well aligned with
the true eigenvectors and lead to dot products of order unity.
These are shown in Figure \ref{fig_cosmicweb}. We find for $[\hat{e}_1,\hat{e}_2,\hat{e}_3]$ average alignment cosine angles of $[0.80,0.70,0.80]$ for \texttt{T-TomoDESI}, $[0.87,0.79,0.80]$ for  \texttt{T-CLA/PFS}, and $[0.96,0.92,0.96]$ for \texttt{T-30+T}. This is again comparable to the results derived from Wiener-filtered flux maps in \cite{2017Krolewski} for the CLAMATO/PFS case, but probing smaller scales.

\begin{figure*}
  \begin{center} 
  \includegraphics[trim=0cm 0cm 0cm 0cm,width=0.90\textwidth]{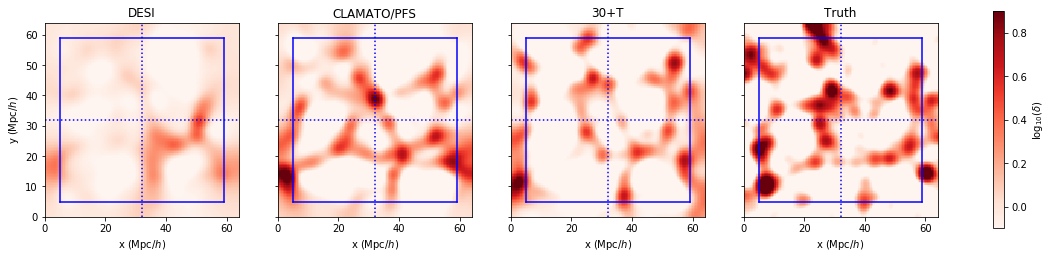}
  \end{center}
  
  \hspace{30.5pt}
  \includegraphics[trim=0cm 0cm 0cm 0cm,width=0.80\textwidth]{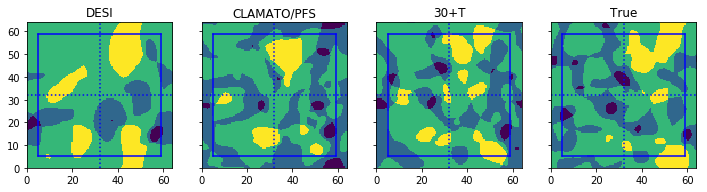}
    \caption{\label{fig_cs_z=0} Comparison of the $z=0$ inferred cosmic structure in Eulerian space, from \texttt{T-TomoDESI}, \texttt{T-CLA/PFS}, and \texttt{T-30+T}, vs. the true z=0 density field. Fields have been smoothed at 2 \mpc. Top: matter density, Bottom: classification of cosmic structure. Dark blue indicated node, light blue indicates filament, green indicated sheet, and yellow indicates void. While the exact location of structures is poorly constrained in real space, the overall structure is quite similar especially with tight sightline spacing.
    }
\end{figure*}
\subsection{Matter Density at $z =0 $}

A main motivation for the TARDIS framework is inferring the late time fate of structures and constituent galaxies found in regions observed by Lyman-$\alpha$ forest tomography. As output of our model, we further evolve the particle field to $z=0$ in order to study the reconstruction. We compare this evolved field with the TreePM `truth' at $z=0$. The true underlying field contains cosmic structures with mass fraction $[0.15, 0.49, 0.31, 0.05]$ and volume fraction $[0.02, 0.28, 0.48, 0.22]$ for [nodes, filaments, sheets, voids], respectively.

Eulerian (real) space provides a qualitative picture of the the structures reconstructed in this limit. In Figure \ref{fig_cs_z=0} (top) we show the matter field and cosmic web 
reconstructed for different survey mock data. While they are qualitatively similar, as described in Subsection \ref{subsec_z=2.5}, the peaks of the $z=2.5$ density field are poorly reconstructed for realistic survey parameters. This results in significant drift of the Eulerian space structures and makes point by point comparisons difficult. This can be seen in Figure \ref{fig_cs_z=0} (bottom) where the qualitative structure is quite similar, especially for 30+T, but the exact positions of nodes and filaments are in slightly different positions relative to the true matter field.
This leads to unsatisfactory cosmic web recovery when evaluated in the same way as $z=2.5$.

However, the reconstructions' cosmic web fidelity at $z=0$ is a somewhat abstract concept since
the Eulerian matter density field is not accessible via any observations. 
Instead, we can evaluate the reconstructed field in Lagrange space, i.e., tracking the 
$z=0$ environments sampled by test particles observed at $z=2.5$. 
Since we expect galaxies to act roughly like test particles in the large-scale
gravitational potential, this provides a direct connection to understanding the late time fate of $z\approx 2.5$ galaxies observed
in the same volume as the Lyman-$\alpha$ tomography data.  
We test this by the following: from the $z=2.5$ density field reconstructed from 
the mock data reconstructions with TARDIS/FastPM, we select a set of test particles
at Eulerian real-space positions $[x_{z25,i}, y_{z25,i}, z_{z25,i}]$ and track them to their $z=0$
Eulerian positions $[x_{z0,i}, y_{z0,i}, z_{z0,i}]$ then evaluate their cosmic web eigenvalues and 
classifications (on the Eulerian real-space grid). From the TreePM `true' matter density field at $z=2.5$, 
we find matching test particles at the same Eulerian positions 
$[x_{z25,i}, y_{z25,i}, z_{z25,i}]$ and again track them to their $z=0$ positions and environments. This process is visualized in Fig~\ref{fig_displacement} where we show the displacement vectors for particles from the reconstructions vs. matched particles from the TreePM simulation evolved to z=0. 

\begin{figure*}
  \centering 
  \includegraphics[trim=1cm 0cm 0cm 0cm,width=0.80\textwidth]{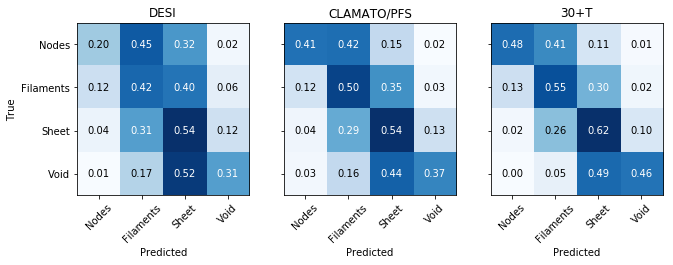}
    \caption{Confusion matrix for cosmic web structures at $z=0$ in Lagrange space (i.e.\ comparing particles with matched in $z=2.5$ positions)
    shown with the reconstructed fraction printed over each cell. While structure is not 
    as well classified as at $z=2.5$, classifications are approximately correct and tend toward morphologically similar environments. For comparison, the mass fraction
    residing in $z=0$ nodes, filaments, sheets, and voids are $[0.15, 0.49, 0.31, 0.05]$, respectively.
    \label{fig_confusion_z=0}}
\end{figure*}

The results from this exercise are shown in the $z=0$ Lagrangian confusion matrix 
in Figure~\ref{fig_confusion_z=0}. For CLAMATO/PFS, we are able to successfully predict the 
$z=0$ environment sampled by the test particles with $\sim 40-50$\% fidelity, while this increases
slightly to $\sim 50-60\%$ in the case of \texttt{T-30+T.} In both cases, $>90\%$ of the particles 
are predicted to lie within $\pm 1$ of the correct cosmic web classification, with the exception of 
CLAMATO/PFS node particles that are misidentified as sheet particles in 15\% of cases.
Nonetheless, this demonstrates the remarkable ability of TARDIS to infer the $z=0$ environment of 
galaxies observed at $z=2.5$, across over 10 Gyrs of cosmic time.

\begin{deluxetable*}{@{\extracolsep{6pt}} l  c c c   c c c c @{}}
\tablecolumns{8}
\tablecaption{\label{tab:cs_z=0} Cosmic Web Recovery at $z=0$ (Lagrangian Comparison) }
\tablehead{ \multirow{2}{*}{Mock Data}
   & \multicolumn{3}{c}{Pearson Coefficients} & \multicolumn{4}{c}{Volume Overlap (\%)} \\ \cline{2-4} \cline{5-8}
    \colhead{}  & \colhead{$\lambda_1$} &  \colhead{$\lambda_2$} & \colhead{$\lambda_3$} & 
      \colhead{Node} & \colhead{Filament} & \colhead{Sheet} & \colhead{Void} }
\startdata
        \texttt{T-TomoDESI} & 0.58 & 0.40 & 0.34 & 20 &  42 & 54 & 31\\
        \texttt{T-CLA/PFS} & 0.70 & 0.54 & 0.47 & 41  & 50 &  54 & 37\\
        \texttt{T-30+T}  & 0.82 & 0.67 & 0.54 & 48  & 55 & 62 & 46\\
\enddata
\end{deluxetable*}

%\subsection{Velocity at $z\sim 2.5$}

%[s/n-type calculation?]

\section{Conclusion}
\label{sec:conclusion}

We present the first use of initial density reconstruction on densely-sampled \lya\ forest data sets (often called ``IGM tomography''), and have showed that 
using this technique we are able to accurately reconstruct large scale properties within the survey volume over a range of scales. In particular, we are able 
to recover the characterization and orientation of the cosmic web at $z=2.5$ in terms of the deformation eigenvalues and eigenvectors assuming
mock data that reflect upcoming and future multiplexed spectroscopic instrumentation. In addition, we are able to recover the qualitative structure of the observed structures at late time, $z=0$. We have also shown that the inferred flux maps from TARDIS are more accurate and have less variance than those from Wiener filtering. Excitingly, we argue that we would be able to \textit{predict the late-time environments of $z\approx 2.5$ galaxies} that
are coeval with our reconstructed IGM tomography volume. This provides a promising and direct route to studying galaxies and AGN in the context of their surrounding cosmic web. For example, we would be able to identify the direct progenitors of $z=0$ filament galaxies, and study their $z=2.5$ galaxy properties.
 While we are currently limited by noise levels and sight-line spacing, in future papers we will explore ways to correct
for underestimated fluctuation amplitude as a function of survey parameters.

While only explored indirectly (through $z=0$ density reconstruction) a direct product of this technique is the particle velocity field at $z=2.5$ which could have significant uses in informing astrophysical processes as well as cosmological constraints. For example, it could allow accurate estimation of velocity 
dispersions in high-redshift
protoclusters, which is currently uncertain due to challenges in disentangling galaxy peculiar motions from the large-scale Hubble expansion \citep{wang2016,topping:2018, cucciati:2018}. More generally, the velocity field reconstruction extends over the entire field and could be a useful addition beyond velocity fields from galaxy redshift space distortions and kinetic Sunyaev Zeldovich effects \citep{2017Sugiyama}. While one might hope to use this reconstruction for constraining other exotic physics (such as using void velocity profiles to provide constraints on modified gravity \citep{2018Falck} and neutrino mass \citep{2015Massara}) the nature of our forward model will restrict the reconstructed maps to obey a $\Lambda$CDM cosmology. If alternative models were implemented efficiently into an $N$-body solver their validity could be tested by comparing the best fit likelihood values.

%Reconstructing the velocity information from \lya was explored in \citet{2001Pichon}, where they perform a joint reconstruction of the matter and velocity field in the observed volume, assuming a statistical correlation between the two fields. 

%\onecolumngrid
%\twocolumngrid

In this work we have held the astrophysical and cosmological parameters constant. A more complete treatment would require varying these jointly with the underlying field; however, we view this as unnecessary at this point since existing data covers a very limited volume with minimal cosmological constraining power. For next generation surveys, which will greatly expand the footprint covered, it will be required to jointly vary these parameters as well. Within the FGPA approximation the astrophysical parameters aren't a significant limitation since there are only a two global parameters of interest ($A_0$,$\gamma$) and our optimization scheme is fast enough that a naive Markov Chain Monte Carlo sampling would be sufficient to explore this parameter space. We explored the sensitivity of the reconstruction with respect to the absorption model in Appendix \ref{app:sens}.

Our focus in this work is on reconstructing the moderate-density large scale structure within the survey volume, and we demonstrated that we were able to recover qualitative structure over a range of scales. Going forward, it would be useful to study how well similar techniques would be on reconstructing halo-scale (i.e.  $\leq 1$ \mpc) structure, such as stacked halo and void profiles.
However, going to this small scale regime reconstruction will be limited by the the specific astrophysical processes within the high-density regions where the Fluctuating Gunn Peterson Approximation will no longer hold. In particular, numerical hydrodynamic simulations have shown that there are significant deviations away from a simple temperature-density scaling relationship close to halos, in some cases even showing a turnover of the relationship \citep{2018Sorini}. It should be possible to extend the formalism proposed in this work and treating the variations from FGPA with some additional parameters to be fit for (or marginalized) in this limit, such as was done for galaxy surveys via a bias expansion \citep{2015ata,2016Kitaura,2018BORG}. One could also use grid-based approximation methods for baryonic effects  \citep[such as][]{2018Dai} to provide a more precise formation formalism for halo substructure, or use a more accurate N-body-based approximation than FGPA \citep{2016Sorini}. It would be a natural extension to test this method on mock data generated from the NyX hydrodynamic simulations designed to accurately reproduce Lyman-$\alpha$ absorption physics \citep{2013nyx,2015nyxlya}. Other non-Tomographic techniques have shown great promise in detecting high redshift clusters from Lyman-$\alpha$ observations \citep{mammoth1}, including a detection of a cluster at $z=2.32$ \citep{mammoth2}, but these techniques probe scales of $\approx 10$ \mpc.

On the other side, additional work is needed to make this reconstruction technique useful for full scale cosmological analysis. Directly extracting power power-spectra estimates from our reconstructed maps suffers from significant noise bias effects which would make them difficult to apply directly to constrain cosmological parameters, as well as mode coupling effects due to the complexity of our forward model. Using a response formalism  \citep[as in][]{seljak2017towards,2018Horowitz,2018fengseljakzaldarriaga} to estimate band-powers would be straightforward and would require $O(N)$ additional optimization runs to estimate $N$ band-powers. However, before using these reconstructions for cosmological analysis, additional considerations are necessary, such as incorporating light-cone effects (i.e. evolution) within the survey volume and including correlated error within our model. While work in this direction is ongoing, upcoming and proposed Lyman Alpha Tomography surveys will cover only a small sky fraction and are unlikely to be directly competitive with other cosmological surveys.

For future reconstruction efforts, the combination of galaxy surveys and Ly$\alpha$ tomographic mapping will be necessary in order to probe different redshift ranges with maximum efficiency.
%Also, as known from seminal works like \citet{bbks}, we know that most massive structures reside at the peaks of the initial densities.
By including the galaxy density field in the reconstruction, we will be able to measure  over-densities with higher precision than from IGM tomography alone. 
Furthermore, incorporating baryonic effects
from hydrodynamical simulations can show how different components of the IGM trace the cosmic web at different redshifts  \citep{Martizzi:2018}. This will allow a joint understanding of the galaxy and IGM large-scale structure distribution and how they influence each other.

\section*{Acknowledgments}
We appreciate helpful discussions with Uro${\rm \check{s}}$ Seljak, Zarija Lukic, Chirag Modi, Teppei Okumura, Yu Feng, and David Spergel.
BH is supported by the NSF Graduate Research Fellowship, award number DGE 1106400, and by JSPS via the GROW program. Kavli IPMU was established by World Premier International Research Center Initiative (WPI), MEXT, Japan. This work was supported by JSPS KAKENHI Grant Number JP18H05868.

This research used resources of the National Energy Research Scientific Computing Center, a DOE Office of Science User Facility supported by the Office of Science of the U.S. Department of Energy under Contract No. DEC02-05CH11231.

\appendix

\section{Convergence}
\label{app:converge}
An important question with any optimization scheme is the convergence properties of the procedure. This is particularly important for nonlinear processes like structure evolution where the likelihood surface is non-Gaussian and conceivably non-convex. We divide the issue into two questions to explore in this appendix; how many iterations are necessary for to be confident in our reconstruction technique and how sensitive is the found solution to the initial optimization starting point? For both questions we explore as a function of scale by looking at the reconstructed transfer function.

\begin{figure}
  \centering  \includegraphics[trim=0cm 0cm 0cm 0cm,width=0.50\textwidth]{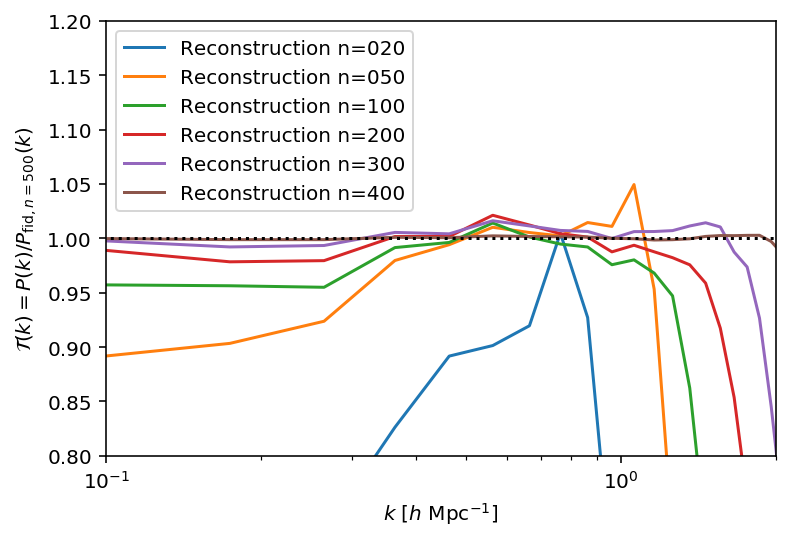}
    \caption{Transfer function with respect to a well converged solution as a function of iteration number. As the iteration number progresses, smaller and smaller scales converge. In addition, there are larger modes on order the box size that are similarly slow to converge.} 
    \label{fig_convergence}
\end{figure}

It has been shown that in the very low noise limit the likelihood surface of possible initial conditions in multi-modial; i.e. gravitational evolution is a non-injective map from initial conditions to late time structure \citep{2018fengseljakzaldarriaga}. However, this uncertainty is due to the shell-crossing degeneracy, which is only relevant for small scale non-linear structure not observed by even the optimistic configurations considered in this work.
%due to spectrographic smoothing, sight-line density and noise levels. 
To study whether or not there is one "true" solution or if there exist sufficiently different converged solutions, we perform the optimization analysis for the same mock catalog with different optimization starting points. In particular, we randomly choose a wide range of initial white noise fields with variance spanning three orders of magnitude. We calculated their transfer functions after 100 iterations versus a fiducial ``well-converged" solution which underwent 500 iteration steps. Up to the scales of interest for the structures studied in this work, $\approx 1$ \mpc\, we find very good agreement between all different starting points. There are some differences of power on very large scales, reflective of the poor constraining power of modes of order the box-size. The number density of modes per uniform bin scales as $k^2$, resulting in significantly more weight placed on smaller modes, until the window function (depending on the smoothing scale and sight-line density) creates a sharp cutoff. If these larger modes are of significant interest, an adiabatic optimization scheme could be used where-in the optimization begins first on a smoothed version of the observed field and then slowly small scale power is introduced back in by varying the smoothing scale as the optimization progresses  \citep[as done in][]{2018fengseljakzaldarriaga}), or potentially directly using a multi-grid preconditioner technique \citep{2007multigrid}. Utilization of these techniques will likely be useful when extending this work for cosmological analysis.

The next important consideration is how long our scheme takes to be fully converged. We plot the transfer function as a function of convergence step in Fig~\ref{fig_convergence}. The exact choice of cutoff depends on the scales of interest, but since we are fundamentally limited in the transverse direction by the line of sight density and in the longitudinal direction by the spectrograph resolution, power above $k = 1.0$ h/Mpc is mostly lost to the smoothing operations on our field. By $n=100$ we find good agreement up to $k = 1.0$ (h/Mpc) and we use this criteria as an iteration limit in the main work.

%\begin{figure}
%  \centering  \includegraphics[trim=0cm 0cm 0cm 0cm,width=0.90\textwidth]{./appendix_figures/different_starts.png}
%    \caption{\texttt{T-CLAMATO} Transfer function of various solutions with different starting points. [Temporary Figure]} 
%    \label{appfig_startingpoints}
%\end{figure}

%\begin{figure}
%  \centering  \includegraphics[trim=0cm 0cm 0cm 0cm,width=0.90\textwidth]{./figs_fastpm/flux_sn.png}
 %   \caption{CLAMATO-like MOCK (FASTPM) S/N over iteration number} 
%    \label{fig_sims2x2}
%\end{figure}

\section{Sensitivity to Cosmology and Absorption Model}
\label{app:sens}

In the main body of this work we have held cosmological and astrophysical parameters constant for the reconstructions. Here we briefly explore how wrong assumptions about the astrophysics or cosmology would bias our late time density field. 

We use a different mock catalog, \texttt{T-IDEAL}, in order to examine the effects of varying the astrophysical parameters. This catalog has a constant signal to noise of $50$ along each skewer, no continuum error, and a sightline density twice that of \texttt{T-30+T}. The idea of this super-experiment is to isolate the effects of the astrophysics from other potential sources of noise in the reconstruction. We perform our reconstructions assuming the ``truth" astrophysics from our mock catalog, as well as assuming the wrong the overall flux amplitude, $A_0 = \exp{(-T_0)}$, and the density scaling exponent, $\beta$.

\begin{figure}
\centering 

  \includegraphics[trim=3cm 0cm 0cm 0cm,width=1.2\textwidth]{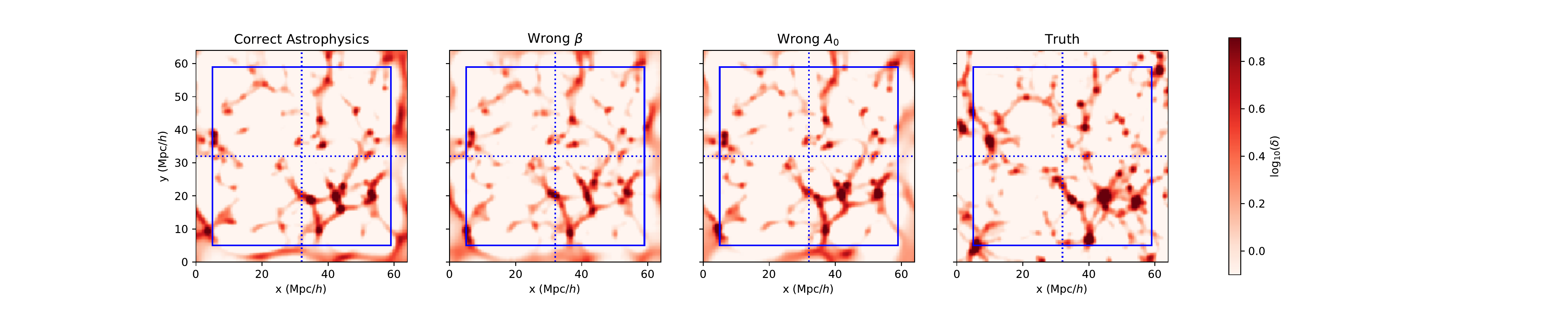}

  \includegraphics[trim=0cm 0cm 0cm 0cm,width=0.70\textwidth]{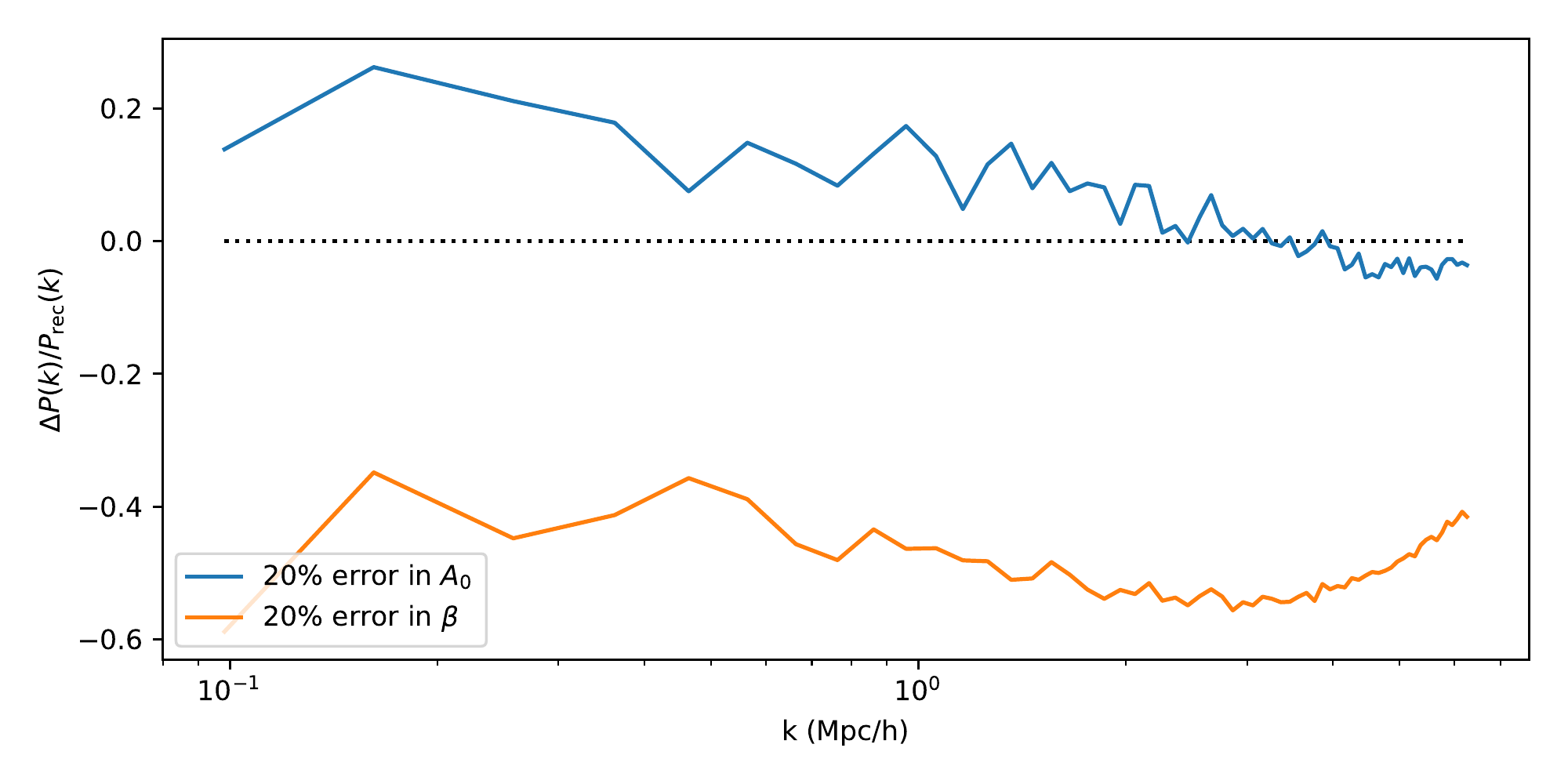}
    \caption{Effect of assuming the wrong astrophysical parameters on the $z=0$ structure, both for a slice in real space (top) and the power spectra (bottom). Even under wrong astrophysical assumptions, we recover similar cosmic structures.} 
    \label{fig_astro}
\end{figure}

We see the effects of wrong astrophysical assumptions in Fig~\ref{fig_astro}. Even with rather radically different astrophysical assumptions we find similar qualitative features in the late time structure. On the power-spectra level, we find these wrong assumptions result primarily in a bias offset from the true power-spectra. In practice, for surveys of the size studied in this work, it would be easily numerically tractable to sample over these parameters to perform the late time reconstruction or, alternatively, to use Lyman Alpha Tomography as a constraint on these parameters.

\section{Comparison to Wiener Filtering}
\label{app:wf}

\begin{figure}

  \centering  \includegraphics[trim=0cm 0cm 0cm 0cm,width=0.80\textwidth]{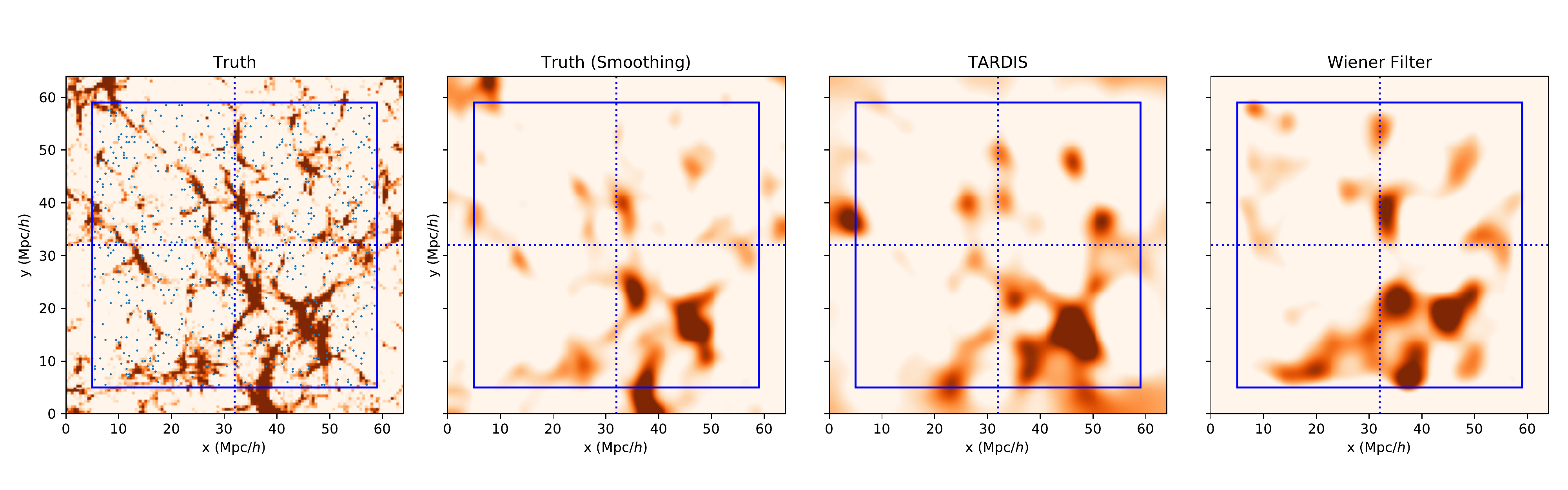}
  \centering  \includegraphics[trim=0cm 0cm 0cm 0cm,width=0.80\textwidth]{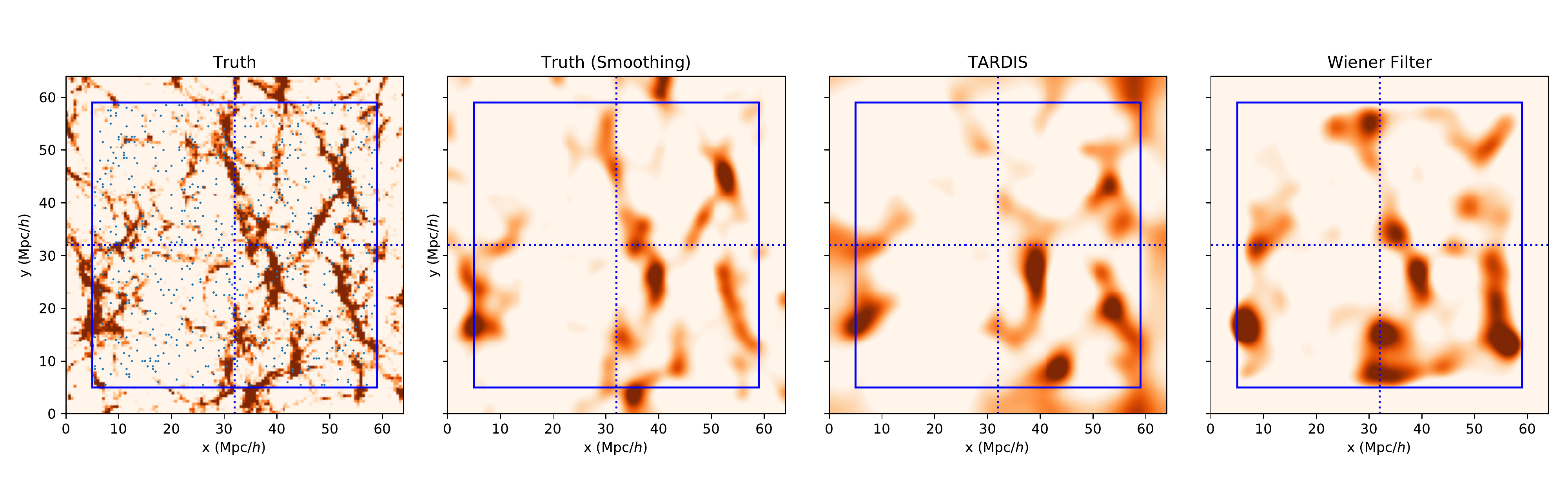}
  \centering  \includegraphics[trim=0cm 0cm 0cm 0cm,width=0.80\textwidth]{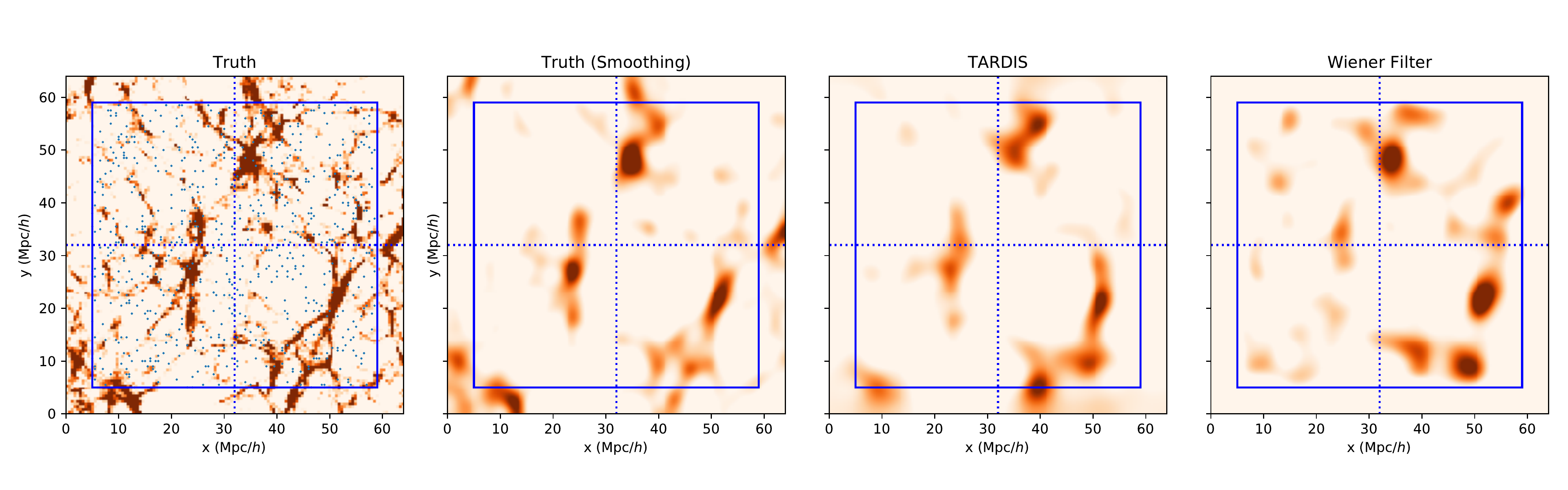}
    \caption{Comparison of the true field, \texttt{TARDIS} reconstructed field, and the Wiener filtered field for the \texttt{T-CLA/PFS} mock. In the far left panels we show the unsmoothed true flux field, with sightlines indicated as blue dots. The blue box indicates boundaries of the survey, with the blue cross to help aid the eye in matching structures. We smooth the 3 rightmost column maps on 2 \mpc and project over a 5 \mpc slice. The recovered flux field is fairly similar between TARDIS and the Wiener filter.} 
    \label{fig_wfcompare}
\end{figure}

A promising aspect of this initial density reconstruction technique is that the reconstructed $z\sim 2$ flux field should be strictly more accurate than that from direct Wiener filtering (WF) of the skewers. This is because direct WF is a purely statistical process which does not take into account the physical evolution of the system under gravity, which further constrains the observed flux field. In this section we review the WF technique which we compare our method against. For a more general discussion of efficient WF and associated optimal bandpower construction, see \citet{seljak1998cosmography} and \citet{2018Horowitz}. For a more through description in the context of Lyman alpha forest, see \citet{Stark2015}.

As we are trying to reconstruct the optimal map given the data, we have to take into account both the data-data covariance, $\textbf{C}_{\textrm{DD}}$, the map-data covariance, $\textbf{C}_{\textrm{MD}}$, and the overall map noise covarance, $\textbf{N}_{ij}$. The reconstructed map can then be expressed in terms of the observed flux, $\delta_F$ as a standard Wiener filter by ;

\begin{equation}
\delta_F^{\textrm{rec}} = \textbf{C}_{\textrm{MD}} \cdot
(\textbf{C}_{\textrm{DD}} + \textbf{N})^{-1} \cdot \delta_F.
\label{eqn:wiener}
\end{equation}
We approximate the covariance by assuming that $\textbf{N}_{ij}
= n_i^2 \delta_{ij}$ where $n_i$ is the pixel noise. This neglects the correlated error component of continuum errors, but this is sub-dominant to the spectrograph noise and shouldn't appreciably affect our reconstructed maps. The map-data and data-data covariances are therefore approximated as 
\begin{equation}
C = \sigma_F^2 \exp{\left[-\frac{\Delta x_{\perp}^2}{2 l_{\perp}^2}
- \frac{\Delta x_{\parallel}^2}{2 l_{\parallel}^2}
\right].}
\label{eqn:covariance}
\end{equation}

In order to compare directly to the Wiener filter map we use the inferred reconstructed flux map from TARDIS.

We apply the Wiener filtering algorithm to the \texttt{T-CLA/PFS} mock catalog and compare along a number of slices to the \texttt{TARDIS} reconstruction. The results are shown in Fig~\ref{fig_wfcompare}. Overall there is good agreement between all maps, with certain smaller-scale features better reconstructed in the \texttt{TARDIS} maps than the Wiener filtered maps.

\begin{figure}
\begin{center}
\begin{overpic}[width=0.480\textwidth]{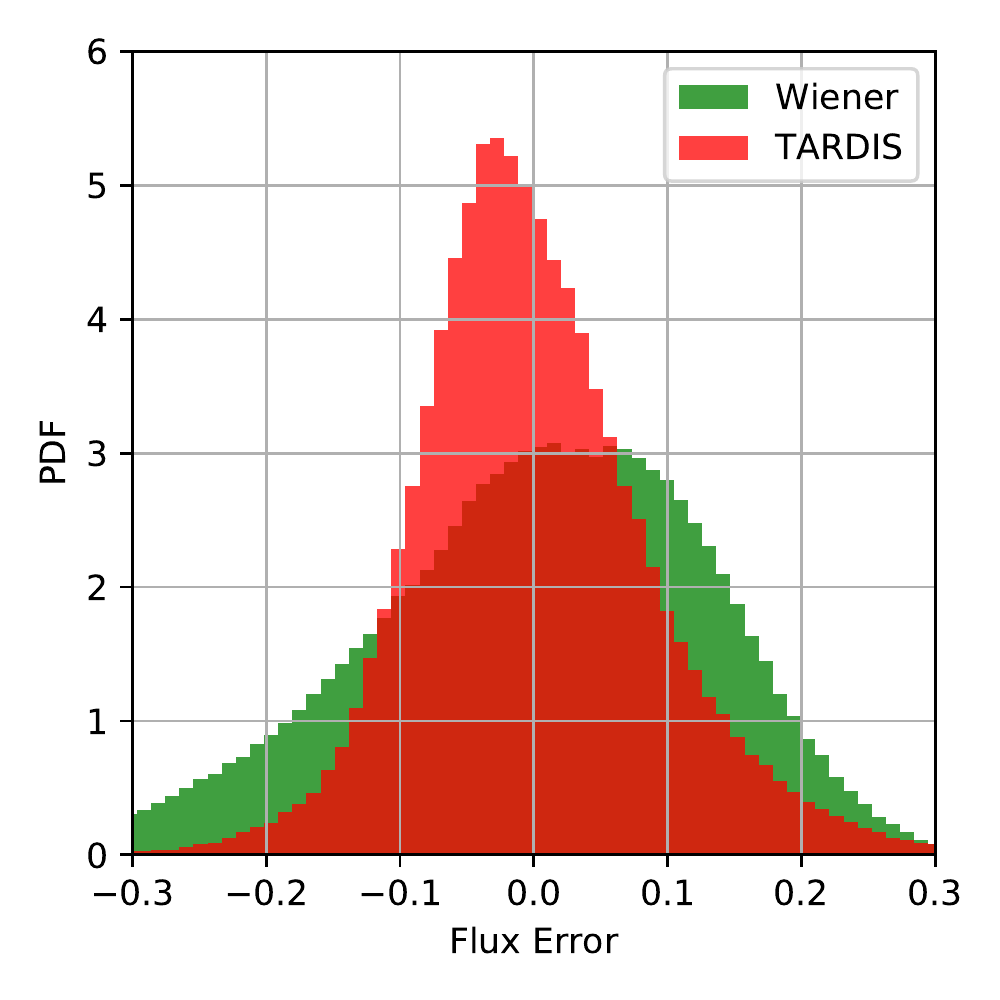}
\put(60,-10){\textsf{\scriptsize (a) Pixel flux error}}
\end{overpic}
\vspace{1em}
\begin{overpic}[clip,trim={0cm 0cm 0 0cm},width=0.495\textwidth]{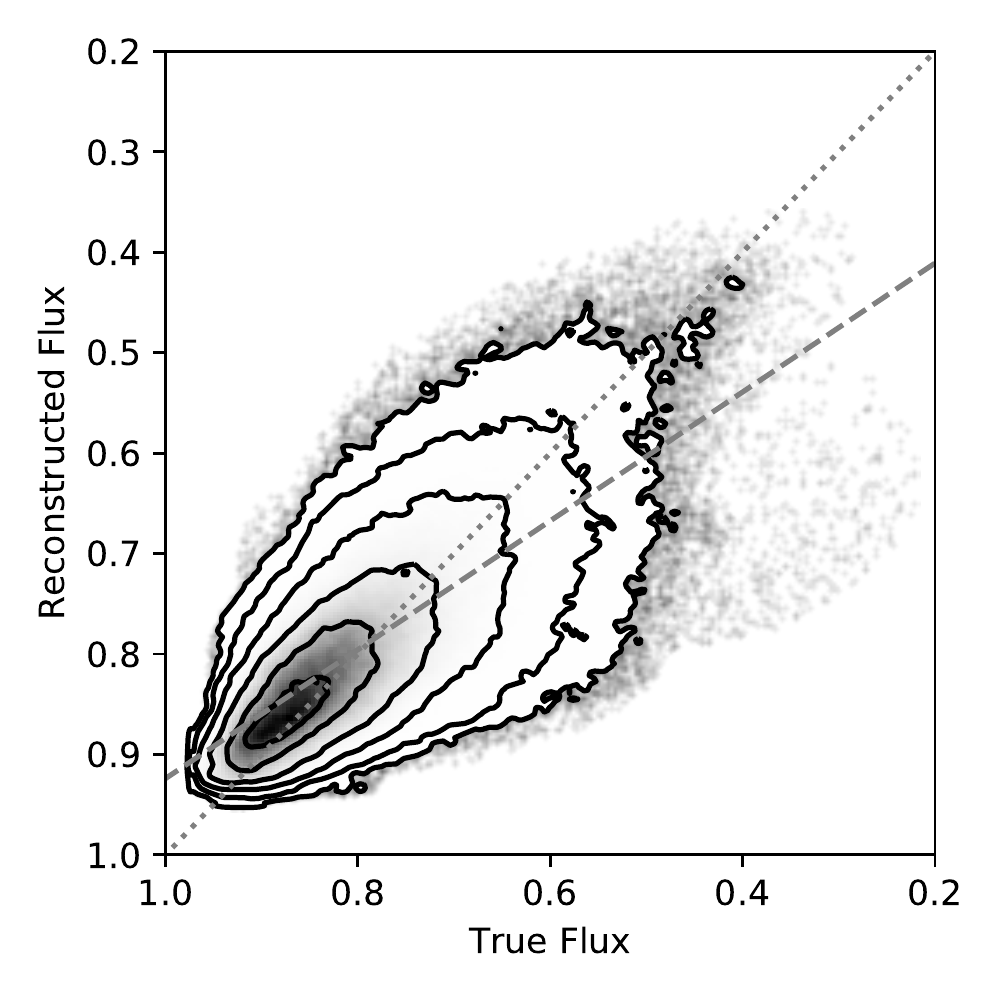}
\put(60,-10){\textsf{\scriptsize (b) Flux reconstruction}}
\end{overpic}
\end{center}

    \caption{Comparing the flux reconstruction for the \texttt{T-CLA/PFS} mock catalog. For these comparisons we have taken a central box which is $35$ \mpc side-length in order to mitigate potential boundary effects and smoothed the region with a  1.5 \mpc Gaussian. In this plot we work in redshift space, unlike the other plots in the paper. $(a)$ Comparison of the corrected fluxes for the Wiener filter map and \texttt{TARDIS} reconstruction vs. the true flux. $(b)$ Scatterplot of the \texttt{TARDIS} reconstructed corrected flux vs the true flux. Also shown is the linear fit of the uncorrected flux (dashed grey line) which was linearly transformed to the $x=y$ dotted line. If interpreted as a flux PDF, each level surface indicates $0.5 \sigma$ density. After this linear correction, the resulting TARDIS flux has no significant bias and mildly outperforms a linearly-corrected Wiener filtered map.} 
    \label{fig_fluxcompare}
\end{figure}

 A well-known feature of reconstructed maps are the presence of a bias caused by the presence of noise. We correct for this bias by a linear transformation calibrated from a separate simulated volume. The effect of this transformation is shown in Fig \ref{fig_fluxcompare} (b). We show the reconstructed flux error in \ref{fig_fluxcompare} (a), showing that the TARDIS maps have smaller flux error variance than the Wiener filtered maps.

\bibliographystyle{apj}

\bibliography{sample}

\end{document}